\documentclass[a4paper,11pt]{article}
\makeatletter
\def\@fpheader{}
\makeatother
\usepackage{jheppub}
\usepackage[utf8]{inputenc}
\usepackage[T1]{fontenc}
\usepackage{ifthen}
\allowdisplaybreaks
\usepackage{amssymb}
\usepackage{amsmath}
\usepackage{bm}
\usepackage{graphics} 
\usepackage{graphicx}
\usepackage{booktabs} 
\usepackage{array}  
\usepackage{xcolor}
\usepackage{caption}
\usepackage{subfig}
\usepackage{graphicx}
\usepackage[compat = 1.1.0]{tikz-feynman}
\usepackage{tikz}
\usepackage{multirow}
\usepackage{adjustbox}
\usepackage{pgfplots}
\usepackage{epsfig}
\usepackage{slashed}
\definecolor{Dblue}{rgb}{0.1, 0.1, 0.8}
\definecolor{Lblue}{rgb}{0.22,0.51,0.9}
\definecolor{green}{RGB}{53,170,102}
\definecolor{Green}{rgb}{0.0, 0.5, 0.0}
\definecolor{Bred}{rgb}{0.8, 0.25, 0.33}
\definecolor{DBlue}{rgb}{0.1, 0.1, 0.8} 
\definecolor{LBlue}{rgb}{0.22,0.51,0.9}
\definecolor{green}{rgb}{0.0, 0.5, 0.0}
\definecolor{BRed}{rgb}{0.8, 0.25, 0.33}
\usepackage{orcidlink}
\usepackage{hyperref}
\usepackage{etoolbox}
\hypersetup{linktocpage,colorlinks=true, citecolor=BRed,linkcolor=green, urlcolor=DBlue,anchorcolor=blue}

\title{\boldmath Mono-Z Signatures of Vector Dark Matter Detection at Future Electron-Positron Colliders: FCC-ee and ILC}
\author[a]{Seyed Yaser Ayazi,\orcidlink{0000-0002-5994-3316}}
\author[b]{ Saeid Paktinat Mehdiabadi \orcidlink{0000-0001-7303-0217}}
\author[a,c]{and Rouzbeh Rouzbehi, \orcidlink{0009-0005-7139-5841}}
\emailAdd{syaser.ayazi@semnan.ac.ir}
\emailAdd{spaktinat@yazd.ac.ir}
\emailAdd{rrouzbehi@chungbuk.ac.kr}
\affiliation[a]{Department of Physics, Semnan University, P.O. Box. 35131-19111, Semnan, Iran}
\affiliation[b]{Department of Physics, Yazd University, P.O. Box 89195-741, Yazd, Iran}
\affiliation[c]{Department of Physics, Chungbuk National University, Cheongju, Chungbuk 28644, Korea}

\DeclareUnicodeCharacter{2212}{-}
\abstract{We investigate the mono-$Z$ signature of a vector dark matter (VDM) model at future electron-positron colliders. In this model, a stable vector boson, arising from an extended $U(1)$ gauge symmetry, serves as a viable DM candidate that interacts with the Standard Model (SM) through a Higgs portal. We perform a detailed cut-based analysis for the mono-$Z$ channel at the FCC-ee (with $\sqrt{s}=240$ and $365$ GeV) and the ILC (with $\sqrt{s}=1$ TeV), considering both leptonic and hadronic decays of the $Z$ boson. Signal significances are evaluated for benchmark points using fast event simulation with MadGraph5, PYTHIA8, and Delphes3. Our results demonstrate that future $e^+e^-$ colliders provide excellent sensitivity to Higgs-portal VDM, with the ILC's beam polarization offering an additional handle to suppress SM backgrounds. This study highlights the potential of lepton colliders in probing DM interactions.}
\keywords{Beyond the Standard Model, Dark matter, Future $e^+e^-$colliders.}

\begin{document}
	\maketitle
	
	\section{Introduction}

	The nature of dark matter (DM) remains one of the most important open questions in particle physics and cosmology; see Refs.~\cite{Cirelli:2024ssz, Bozorgnia:2024pwk} for recent reviews. Within the standard cosmological framework, DM accounts for approximately 27\% of the total energy density of the Universe, with a cold DM abundance of $\Omega_{\rm DM}h^2\simeq 0.120$~\cite{Planck:2018}. Despite the success of the Standard Model (SM) in describing elementary particles and their interactions, it contains no viable candidate for this component. Identifying the particle nature of DM therefore remains a central motivation for physics beyond the SM.
	
	The search for DM relies on complementary approaches involving direct detection, indirect detection, and collider experiments. Collider searches provide an opportunity to produce DM particles in a controlled environment and investigate their interactions with SM particles. Their production rates do not depend on the local DM density or velocity distribution, and their interpretation does not require assumptions about the Galactic halo profile. If DM particles are produced in association with a visible object, their escape from the detector leads to an imbalance in the measured momentum. Such signatures allow collider experiments to constrain the interactions connecting the visible and dark sectors, including processes mediated by additional particles that may also be produced on shell. Collider measurements thus provide information complementary to the scattering and annihilation processes probed by direct and indirect searches.

	Among the many proposed scenarios, DM models based on an extended gauge sector offer a theoretically well motivated framework. In particular, vector DM (VDM)--A stable vector boson from a spontaneously broken $U(1)_D$ gauge symmetry can serve as VDM--has gained attention due to its minimality and testability. The VDM paradigm has been extensively studied in the context of hidden sectors and dark gauge symmetries~\cite{Hambye:2008bq, Lebedev:2011iq, Ko:2014gha, Baek:2012se, Duch:2015jta, Farzan:2012hh}. In this work, we consider a VDM model where the dark sector communicates with the SM particles exclusively through a Higgs portal interaction, with a $Z_2$ symmetry ensuring DM stability. The model introduces only three new free parameters: the dark gauge coupling $g_v$, the VDM mass $M_V$, and the additional scalar mass $M_{H_2}$. This framework is fully developed in our previous works~\cite{YaserAyazi:2019psw,Ayazi:2023, Ayazi:2024}. Other important studies on VDM and related Higgs portal scenarios have been carried out in Refs.~\cite{Chen:2026xov, Chao:2014ina, Hashino:2018zsi, YaserAyazi:2019caf, Djouadi:2011aa, Baek:2014jga, Glaus:2019itb, Amiri:2022cbv, Beniwal:2015sdl, Shaikh:2026ixs, Mohamadnejad:2019vzg, Hosseini:2023qwu, Hosseini:2024ahe, Hosseini:2026peb,Abkenar:2024ket}.
	
	Future electron-positron colliders, such as the FCC-ee and the ILC, provide an exceptionally clean environment for probing such weakly interacting dark sector particles~\cite{FCC:2018evy, ILC:2019gyn}. Unlike hadron colliders, lepton colliders feature well-defined initial states, negligible QCD backgrounds, and precise missing energy measurements. These advantages make mono-$X$ searches (with $X = Z, \gamma, W, h, j$) powerful channels for discovering or constraining DM production. The FCC-ee, as part of the Future Circular Collider study at CERN, is designed to operate at center-of-mass energies ranging from $90$ to $365$ GeV with unprecedented luminosity, reaching integrated luminosities of up to $5$ ab$^{-1}$ at $240$ GeV, and $1.5$ ab$^{-1}$ at $365$ GeV. The ILC, proposed to be built in Japan, is planned to operate at $\sqrt{s}=250$ GeV, $500$ GeV, and $1$ TeV, with the capability of longitudinally polarized electron and positron beams, providing a powerful handle to suppress SM backgrounds and enhance signal sensitivity.	 Among the possible missing-energy signatures, the mono-$Z$ channel is particularly well motivated~\cite{Wan:2014rhl, Yu:2014ula, Dutta:2017ljq, Grzadkowski:2020frj, Ghosh:2026xkw, Kundu:2023}. A further motivation is the interplay with Higgs precision physics. At $e^+e^-$ colliders the process $e^+e^-\to ZH_1$ allows the Higgs production rate to be measured model-independently from the mass recoiling against the $Z$ boson, $M_{\rm rec}^2=s-2\sqrt{s}\,E_Z+m_Z^2$, irrespective of the Higgs decay mode, and hence also gives direct access to invisible Higgs decays~\cite{Abada:2019,Baer:2013}. The mono-$Z$ search studied here is the natural generalisation of this measurement to the case where the recoil system is the pair $VV$ produced by $H_1$, by the second scalar $H_2$, or through the non-resonant continuum.

	In this paper, we focus on the mono-$Z$ signature: $e^+e^- \to Z VV$ where the $Z$ boson decays leptonically ($Z \to \ell^+\ell^-$ with $\ell = e, \mu$) or hadronically ($Z \to jj$), and the VDM particles $V$ escape detection, resulting in missing energy and momentum. We perform a detailed cut-based analysis for the FCC-ee at center-of-mass energies $\sqrt{s} = 240$ and $365$ GeV, as well as for the ILC at $1$ TeV, including the impact of beam polarization. Signal and background events are generated using \textit{MadGraph5}~\cite{Alwall:2014}, showered and hadronized with \textit{PYTHIA8}~\cite{Bierlich:2022}, and passed through a fast detector simulation with \textit{Delphes3}~\cite{deFavereau:2013fsa} using the \textit{IDEA}~\cite{IDEAStudyGroup:2025gbt} and \textit{ILCgen} detector configuration cards. Signal significances are evaluated for benchmark points with VDM masses ranging from $30$ to $250$ GeV and heavy scalar masses from $90$ to $600$ GeV. Our results demonstrate the sensitivity of future lepton colliders to VDM in the Higgs portal framework.

	The remainder of this paper is organized as follows. In Section~2, we briefly describe the VDM model, including its Lagrangian, scalar potential, and the relevant free parameters, as well as theoretical and experimental constraints. Section~3 is devoted to DM searches at lepton colliders, with a detailed cut-based analysis for the mono-$Z$ channel at the FCC-ee. This section is divided into the leptonic mode and the hadronic mode, including event generation, selection cuts, and signal significance calculations for two center-of-mass energies. Section~4 extends the analysis to the ILC at $\sqrt{s}=1$ TeV, highlighting the impact of beam polarization on enhancing signal sensitivity and suppressing SM backgrounds. Finally, Section~5 presents our conclusions. 	
	
	\section{The Model}\label{Model}
	
	Here, we briefly introduce the main sectors of the model, which is fully developed in Refs. \cite{Ayazi:2023,Ayazi:2024}. The model extends the SM by a dark sector in which the $U_D(1)$ Abelian gauge symmetry is spontaneously broken via the dark Higgs mechanism. A complex scalar field $S$, which carries a unit charge under $U_D(1)$, and the $U_D(1)$ gauge field $V_{\mu}$ are introduced. Additionally, we suppose a $Z_2$ symmetry under which the vector field $V_{\mu}$ and the scalar field $S$ transform as follows:
	
	\begin{equation}
		V_{\mu} \rightarrow -V_{\mu}, \quad S \rightarrow S^*
	\end{equation}
	
	To ensure the stability of the DM candidate, we assume that any kinetic mixing between the vector field $V_{\mu}$ and the SM $U(1)$ gauge boson $B_{\mu}$ (i.e., terms like $V_{\mu\nu} B^{\mu\nu}$) is forbidden. Thus, the vector field $V_{\mu}$ serves as a viable DM candidate. The total Lagrangian of the model is given by:
	
	\begin{equation}
		\mathcal{L} = \mathcal{L}_{SM} - \frac{1}{4} V_{\mu \nu} V^{\mu \nu} + (D'_{\mu} S)^* (D'^{\mu} S) - V(H, S),
	\end{equation}
	where $ \mathcal{L}_{SM} $ is the Standard Model Lagrangian excluding the Higgs potential term. The covariant derivative and field strength of $V_{\mu}$ are defined as:
	
	\begin{align}
		D'_{\mu} S &= (\partial_{\mu} + i g_v V_{\mu}) S, \nonumber \\
		V_{\mu \nu} &= \partial_{\mu} V_{\nu} - \partial_{\nu} V_{\mu}.
	\end{align}
	
	The scalar potential, which is renormalizable and invariant under both gauge and $Z_2$ symmetries, takes the form:
	
	\begin{equation}
		V(H,S) = -\mu_H^2 H^{\dagger} H - \mu_S^2 S^* S + \lambda_H (H^{\dagger} H)^2 + \lambda_S (S^* S)^2 + \lambda_{SH} (S^* S)(H^{\dagger} H).
	\end{equation}
	
	The quartic interaction $ \lambda_{SH} (S^* S)(H^{\dagger} H) $ represents the only interaction between the dark sector and the SM. After electroweak symmetry breaking \cite{Ayazi:2023,Ayazi:2024}, the parameters of the model are given by:
	
	\begin{align}
		\nu_2 &= \frac{M_V}{g_v}, \nonumber \\
		\sin \alpha &= \frac{\nu_1}{\sqrt{\nu_1^2 + \nu_2^2}}, \nonumber \\
		\lambda_H &= \frac{\cos^2 \alpha M_{H_1}^2 + \sin^2 \alpha M_{H_2}^2}{2 \nu_1^2}, \nonumber \\
		\lambda_S &= \frac{\sin^2 \alpha M_{H_1}^2 + \cos^2 \alpha M_{H_2}^2}{2 \nu_2^2}, \nonumber \\
		\lambda_{SH} &= \frac{(M_{H_2}^2 - M_{H_1}^2) \sin \alpha \cos \alpha}{\nu_1 \nu_2}.
	\end{align}
	
	Here, $\alpha$ is the mixing angle between weak eigenstates and mass eigenstates, and $\nu_1$ and $\nu_2$ are the vacuum expectation values (vevs) of the Higgs fields. The mass eigenstates of the Higgs fields can be written as:
	
	\begin{align}
		M^2_{H_{2}, H_{1}} = \lambda_H \nu_1^2 + \lambda_S \nu_2^2 \pm \sqrt{(\lambda_H \nu_1^2 - \lambda_S \nu_2^2)^2 + \lambda_{SH}^2 \nu_1^2 \nu_2^2},
	\end{align}
	where we assume $ M_{H_1} = 125 \, \text{GeV}$ and $ \nu_1 = 246 \, \text{GeV}$. It is noteworthy that this construction is minimal, introducing only three new free parameters: the coupling $g_v$ and the mass parameters $M_{H_2}$ and $M_V$.

	Several remarks on the model are useful for the collider analysis below. (i)~The transformation above acts as a dark charge conjugation: it forbids the kinetic mixing $V_{\mu\nu}B^{\mu\nu}$ and any interaction with an odd number of $V$ fields, so that $V$ is stable and is always produced in pairs, as in $e^+e^-\to Z\,VV$. (ii)~After spontaneous symmetry breaking the vector acquires the mass $M_V=g_v\nu_2$, which is the origin of the relation $\nu_2=M_V/g_v$ above; the Goldstone mode becomes its longitudinal polarisation. (iii)~The physical scalars $H_1$ and $H_2$ are mixtures of the doublet and singlet components. In a generic mixing scenario, their couplings to SM particles are the SM ones rescaled by $\cos\alpha$ and $\sin\alpha$, respectively (up to convention), while their couplings to the dark vector are proportional to $g_vM_V$ times the complementary mixing factor. 	
	
	Furthermore, we examine both theoretical and experimental constraints on the parameter space of the model. Theoretical constraints include conditions such as perturbativity, unitarity, and positivity of the potential, as discussed in \cite{Ayazi:2023}. Experimental constraints include relic density measurements and direct detection bounds. 	
	\section{DM Searches at Lepton Colliders}
	Lepton colliders provide a particularly clean experimental environment for probing DM particles, offering distinct advantages over hadron colliders. Unlike hadron machines, where large quantum chromodynamics (QCD) backgrounds, proton remnants, and pile-up effects obscure missing energy signatures, lepton colliders feature well-defined initial states, negligible QCD activity, and essentially no pile-up. These favorable conditions enable precise measurements of missing energy, an observable largely inaccessible at hadron colliders, as well as missing transverse momentum, both of which are key kinematic variables in DM searches. Moreover, the low-background environment facilitates model-independent mono-X searches (with X = $\gamma$, Z, W, h, or j), wherein DM particles are produced in association with a visible recoil object. These channels significantly improve sensitivity to weakly interacting dark-sector particles, establishing lepton colliders as a powerful complement to hadron colliders in the global DM search program. In this section, we focus on DM searches at future lepton colliders, specifically the FCC-ee and the ILC, via the mono-Z channel. Several studies have explored lepton collider signatures of DM~\cite{Barman:2021hhg, Ge:2023wye, Kundu:2021cmo, Fox:2011fx, Yu:2013aca, Essig:2013vha, Yu:2014ula, Freitas:2014jla, Dutta:2017ljq, Horigome:2021qof, Bhattacharya:2022qck, Ma:2022cto, Dong:2025exu, BrahimAit-Ouazghour:2026stq, Liu:2025ori, Buttazzo:2018qqp, Buttazzo:2015bka, Etesami:2021hex, Altmannshofer:2025nbp, Jahedi:2024kvi, Altmannshofer:2023tsa, Barik:2023bgx, Calibbi:2021pyh, Haghighat:2022qyh, Chae:2012bq, Birkedal:2004xn, Konar:2009ae, Bartels:2012ex, Dreiner:2012xm, Liu:2019ogn, Habermehl:2020njb, Kalinowski:2021tyr, Grzadkowski:2020frj, Kriewald:2025bui, Ghosh:2026xkw, Vignaroli:2026gnp, Giffin:2020jtl}.

	\paragraph{Kinematic variables.} The initial four-momentum is known at an $e^+e^-$ collider, so the invariant mass of the invisible system recoiling against a visible system of energy $E_{\rm vis}$ and invariant mass $m_{\rm vis}$ can be reconstructed event by event,
	\begin{equation}
		M_{\rm rec}^2=(p_{e^+}+p_{e^-}-p_{\rm vis})^2=s-2\sqrt{s}\,E_{\rm vis}+m_{\rm vis}^2=2\sqrt{s}\,\slashed{E}-s+m_{\rm vis}^2 ,
	\end{equation}
	where $\slashed{E}=\sqrt{s}-E_{\rm vis}$ is the missing energy used in the cuts below. For the signal, $M_{\rm rec}$ is the invariant mass of the $VV$ pair: it shows resonant peaks at $M_{H_1}$ and $M_{H_2}$ (when $M_{H_i}>2M_V$) on top of a non-resonant continuum, whereas for $ZZ\to\nu\bar\nu\ell^+\ell^-$ it peaks at $m_Z$ and for $WW$ it is broad. For a given visible mass, a cut on $\slashed{E}$ is therefore equivalent to a cut on $M_{\rm rec}$.

	\subsection{Mono-Z Channel at the FCC-ee at $\sqrt{s}=240, 365$ GeV}
	Here, we briefly explain our simulation procedure for generating signal and background events in the mono-Z channel at $e^+ e^-$ colliders. Based on the decay modes of the Z boson, we consider leptonic Z decays into a charged-lepton pair and hadronic Z decays into a quark-antiquark pair reconstructed as two jets, where the Z boson decays into two charged leptons and two jets, respectively. For the signal simulation, we generated the UFO model file by using LanHEP~\cite{Semenov:2014rea}, and then generated 500000 events using \textit{MadGraph5}~\cite{Alwall:2014} for both signal and backgrounds. \textit{PYTHIA8}~\cite{Bierlich:2022} is used to perform parton showering, hadronization, and decay processes. A fast detector simulation of these events is done using \textit{Delphes3}~\cite{deFavereau:2013fsa} with the \textit{IDEA} configuration card~\cite{IDEAStudyGroup:2025gbt}. We also perform our analysis at the FCC-ee center-of-mass energies $\sqrt{s}$ = 240 and 365 GeV, assuming integrated luminosity of $\mathcal{L}_{\text{int}}$ = 5000 and 1500 $\text{fb}^{-1}$, respectively.

	\subsection{Leptonic mode}
\begin{figure}[]
		\centering
		\begin{tikzpicture}
			\begin{feynman}
				\vertex (a);
				\vertex [below left=of a] (b) {$e^-$};
				\vertex [above left=of a] (c) {$e^+$};
				\vertex [right=of a] (d);
				\vertex [above right=of d] (e);
				\vertex [below right=of d] (h);
				\vertex [right=of h] (j) {$V$};
				\vertex [below right=of h] (i) {$\bar{V}$};
				\vertex[above right=of e] (k) {$f$};
				\vertex[below right=of e] (l) {$\bar{f}$};
				\diagram{
					(b) -- [fermion] (a) -- [fermion] (c);
					(a) -- [boson, edge label=$Z$] (d);
					(d) --[boson, edge label=$Z$] (e);
					(d) -- [scalar, edge label=$\rm H_1/H_2$] (h);
					(i) -- [fermion] (h) -- [fermion] (j);
					(l) -- [fermion] (e) --[fermion] (k);
				};
			\end{feynman}
		\end{tikzpicture}
		\caption{Feynman diagrams for DM production for the mono-Z channel. Depending on the decay mode of the Z boson, f can be leptons ($\ell^\pm \equiv e^\pm, \mu^\pm$) or jets ($j \equiv u, c, d, s, b$).}
		\label{diag.XVV}
	\end{figure}
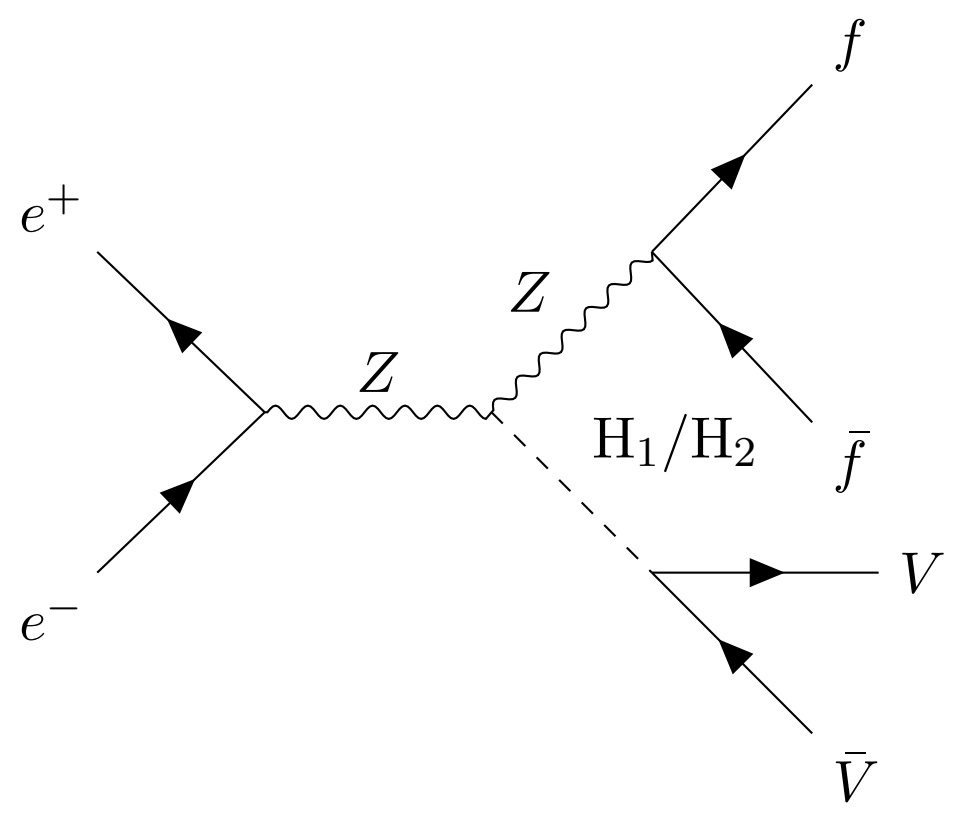

	For the charged leptonic decay of the Z boson, the process $e^+ e^- \to VVZ, (Z\to \ell^+\ell^-)$ is considered. Its tree-level Feynman diagram is shown in Fig.~\ref{diag.XVV}. 
	Only $\ell = e, \mu$ are considered, and $\tau$-lepton is not included for simplicity. For the signal production of VVZ, the final state looks like $\ell\ell + \slashed{E}$, where the missing energy ($\slashed{E}$) is carried away by VDM particles. The main SM background for this channel is $e^+e^- \to \ell^+\ell^- \nu\bar{\nu}$, which comes from the processes of $e^+e^-\to W^+ W^-$ and $e^+e^-\to ZZ$, with the leptonic decays of the vector bosons.

	Subleading SM sources of $\ell\ell+\slashed{E}$ are $e^+e^-\to\tau^+\tau^-$ with leptonic $\tau$ decays, which gives softer lepton pairs that do not peak at $m_Z$ (strongly reduced by Cut~2), the radiative-return process $e^+e^-\to\ell^+\ell^-\gamma$ with the photon lost along the beam pipe (see the kinematic argument below), and two-photon processes, which carry little missing energy (reduced by Cuts~3 and~4).

	\subsubsection{Cut Based Analysis}
	To enhance the signal-to-background ratio, a cut-based analysis is performed. Both signal and background events are generated using a baseline selection: $P_{T_\ell} > $ 2 GeV and $|\eta_\ell| < $ 3.5. The selection cuts for two collision energies are summarized in Tab.~\ref{tab.diLepcut}.
	\begin{table}
		\centering
		\caption{Cross sections $\sigma$ (in fb) and signal significances $\bm{S}$ after all cuts in the charged leptonic channel at $\sqrt {s}= 240$ GeV are shown. The significances are computed by assuming an integrated luminosity of $5~ab^{-1}$. The values in the parentheses denote the significances with 5\% background systematic uncertainty. The value of $g_v$ = 0.1.}
		\begin{tabular}{ccccc}
			\hline
			\hline
			&$M_V$(GeV)&$M_{H_2}$(GeV)&$\sigma$(fb)&$\bm{S}$ \\\hline
			BP1&30&90&160.3&36.20 (1.50)\\\hline
			BP2&50&140&21&43.25 (4.48)\\\hline
			BP3&60&170&5&12.68 (1.22)\\\hline
			$\nu_\ell \bar{\nu_\ell} \ell^+ \ell^- $&&&983.7\\\hline
			\hline
		\end{tabular}
		\label{tab.bps240}
	\end{table}
	
	In the following, a detailed discussion of the selection cuts is provided.
	\begin{itemize}
		\item{Cut 1: Events are requested to contain two opposite-sign, same-flavor (OSSF) leptons, each with $P_{T_\ell} > $ 10 GeV and $|\eta_\ell| < $ 2.8. Events are vetoed if they contain any additional hard lepton, tau lepton, photon, or hadronic jet with $P_T > $ 10 GeV and $|\eta| < $ 2.8. This acceptance cut selects potential events arising from the leptonic decay of the $Z$ bosons.}
		
		\item{Cut 2: The invariant mass of the selected OSSF leptons ($M_{\ell\ell}$) must be consistent with the leptonic decay of a $Z$ boson: $|m_{\ell\ell}-m_Z|\leq $10 GeV.}
		
		\item{Cut 3: The events are asked to have the missing transverse energy, $\slashed{E}_T >$ 10 GeV for $\sqrt {s}=$ 240 GeV and $\slashed{E}_T >$ 15 GeV for $\sqrt{s} =$ 365 GeV. The $\slashed{E}_T$ defined as:
			\begin{equation}	
				\slashed{E}_T = \sqrt{(\sum_{\text{visible}} p_x)^2 + (\sum_{\text{visible}}p_y)^2}. \nonumber
			\end{equation}
			where $p_x$ and $p_y$ are the x and y components of the momenta of all visible final-state particles.}
		
		\item{Cut 4: The last variable is the missing energy ($\slashed{E}$), which is reconstructed using the energy of the initial and final objects. The missing energy is defined as:
			\begin{equation}
				\slashed{E} = \sqrt{s} - \sum_{\text{visible}} E. \nonumber
			\end{equation}
			where E is the energy of each visible final-state particle, and $\sqrt{s}$ is the center-of-mass energy of the collision.
		}
		
	\end{itemize}

	For the leptonic mode, the thresholds on $\slashed{E}$ can be translated into recoil-mass thresholds with $M_{\rm rec}^2=2\sqrt{s}\,\slashed{E}-s+m_{\ell\ell}^2$, which helps to interpret them physically. Setting $m_{\ell\ell}=m_Z$, the thresholds of Tab.~\ref{tab.diLepcut}, $\slashed{E}>140$~GeV at $\sqrt{s}=240$~GeV and $\slashed{E}>190$~GeV at $\sqrt{s}=365$~GeV, correspond to $M_{\rm rec}\gtrsim134$~GeV and $M_{\rm rec}\gtrsim117$~GeV, respectively. At 240~GeV the selection therefore retains the $H_2$ resonance of BP2 and BP3 (140 and 170~GeV) but removes the $H_1$ peak at 125~GeV and the 90~GeV resonance of BP1; at 365~GeV it also retains the $H_1$ peak but still removes the resonance of BP1. This is consistent with the much lower signal efficiencies of BP1 than of BP2 and BP3 in Tabs.~\ref{tab.cutflow240Lep} and~\ref{tab.cutflow365Lep}, and suggests that a benchmark-dependent threshold or a fit of the full $M_{\rm rec}$ distribution could improve the reach for BP1.
	In addition, both thresholds exceed the kinematic endpoint $E_\gamma^{\max}=(s-m_Z^2)/(2\sqrt{s})\simeq103$~GeV (171~GeV) at $\sqrt{s}=240$ (365)~GeV of radiative-return events $e^+e^-\to Z\gamma$ in which the photon escapes along the beam pipe, so this reducible background is removed by Cut~4 in the leptonic mode.
	
\begin{figure}
		\centering
		\subfloat[$\slashed{E}_T$]{\includegraphics[width=0.48\linewidth]{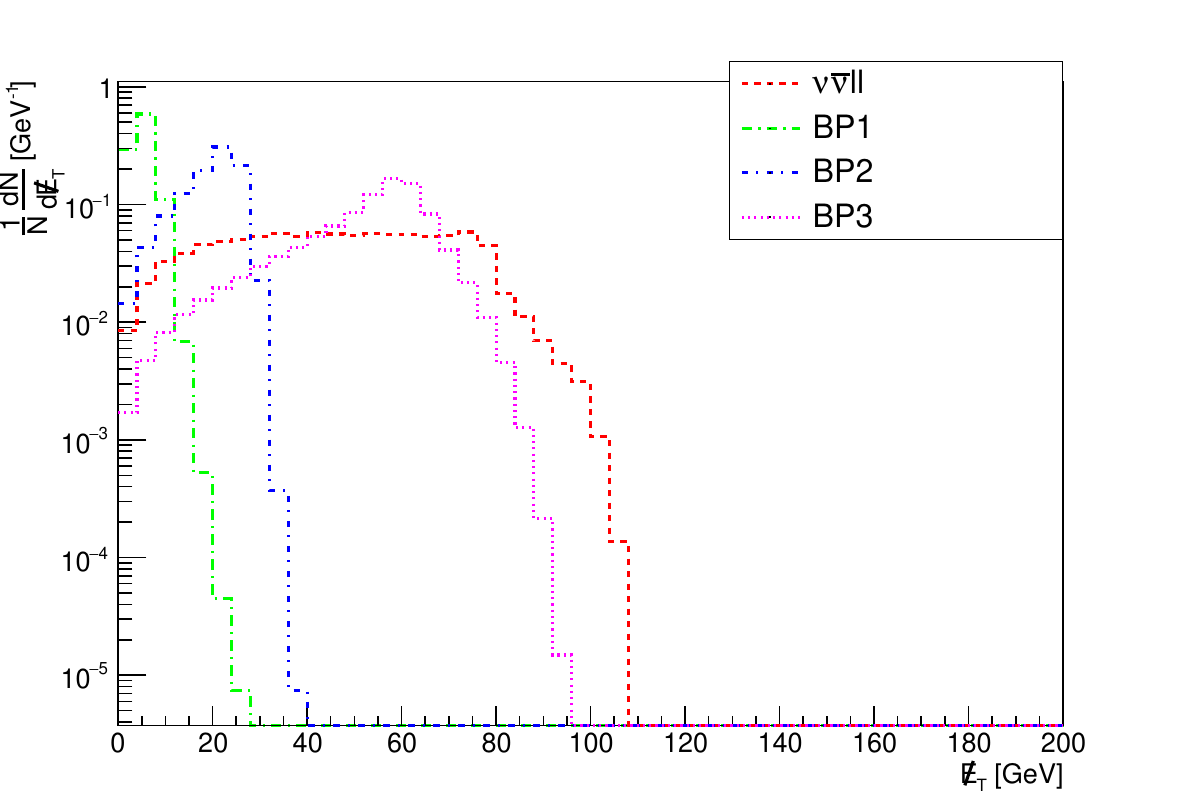}}
		\subfloat[$\slashed{E}_T$]{\includegraphics[width=0.48\linewidth]{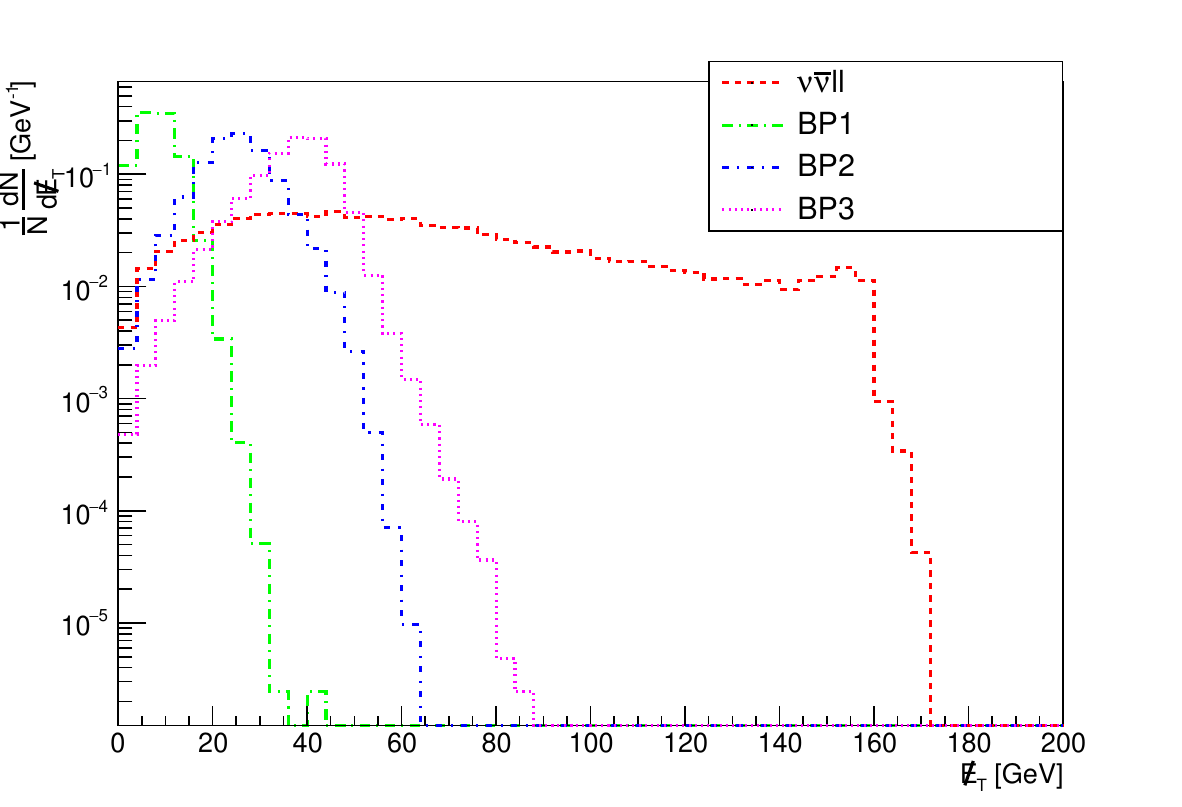}}\\
		\subfloat[M$_{ll}$]{\includegraphics[width=0.48\linewidth]{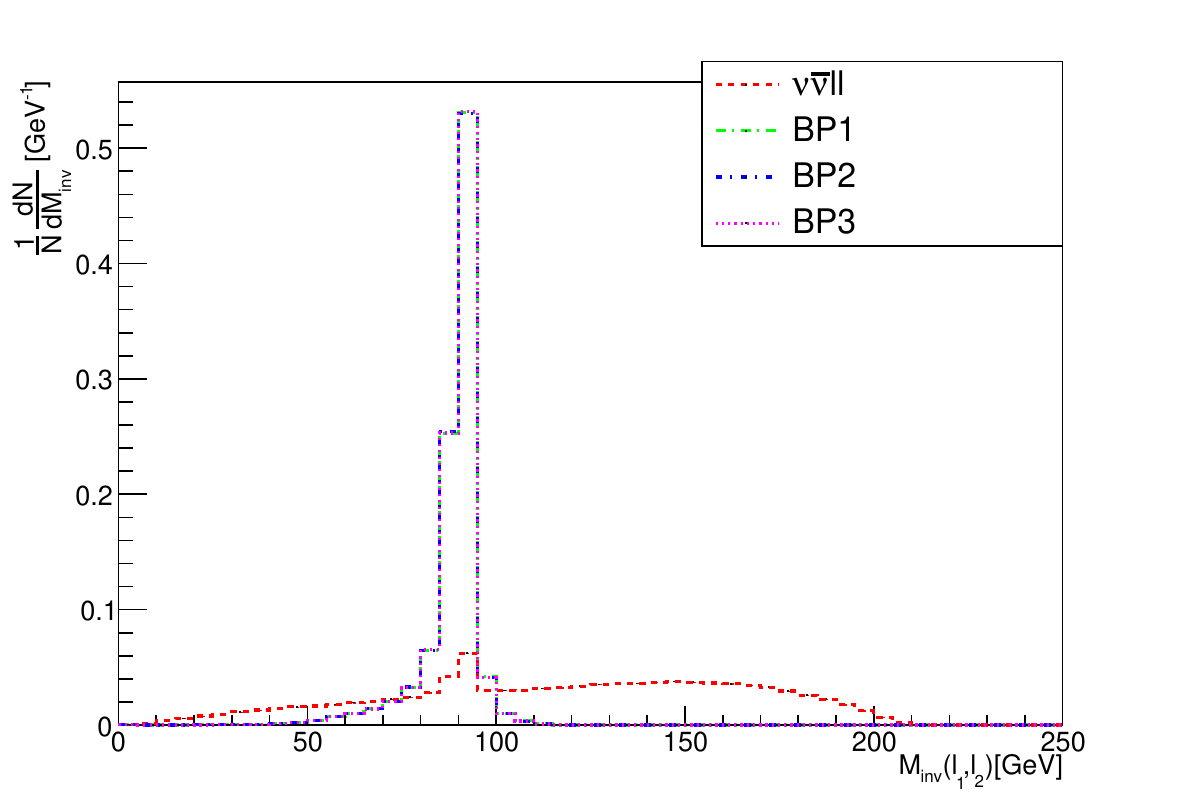}}
		\subfloat[M$_{ll}$]{\includegraphics[width=0.48\linewidth]{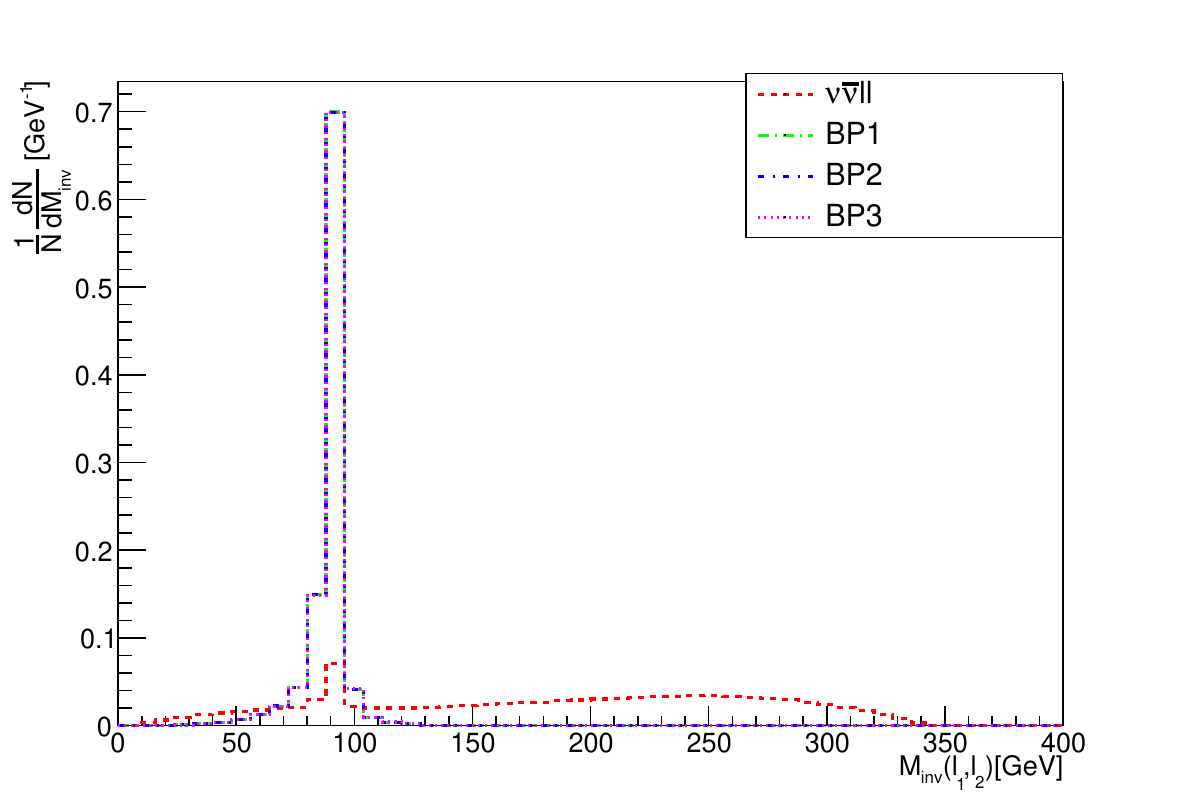}}\\
		\subfloat[MissE]{\includegraphics[width=0.48\linewidth]{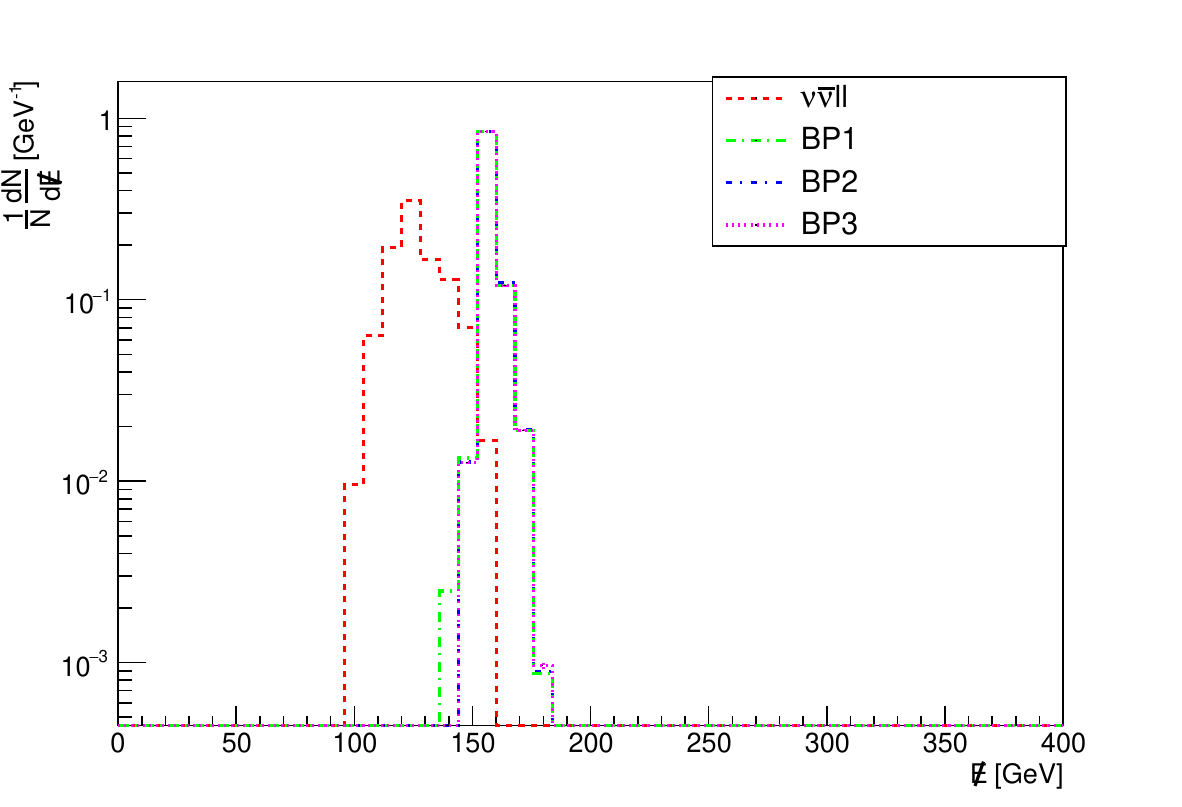}}
		\subfloat[MissE]{\includegraphics[width=0.48\linewidth]{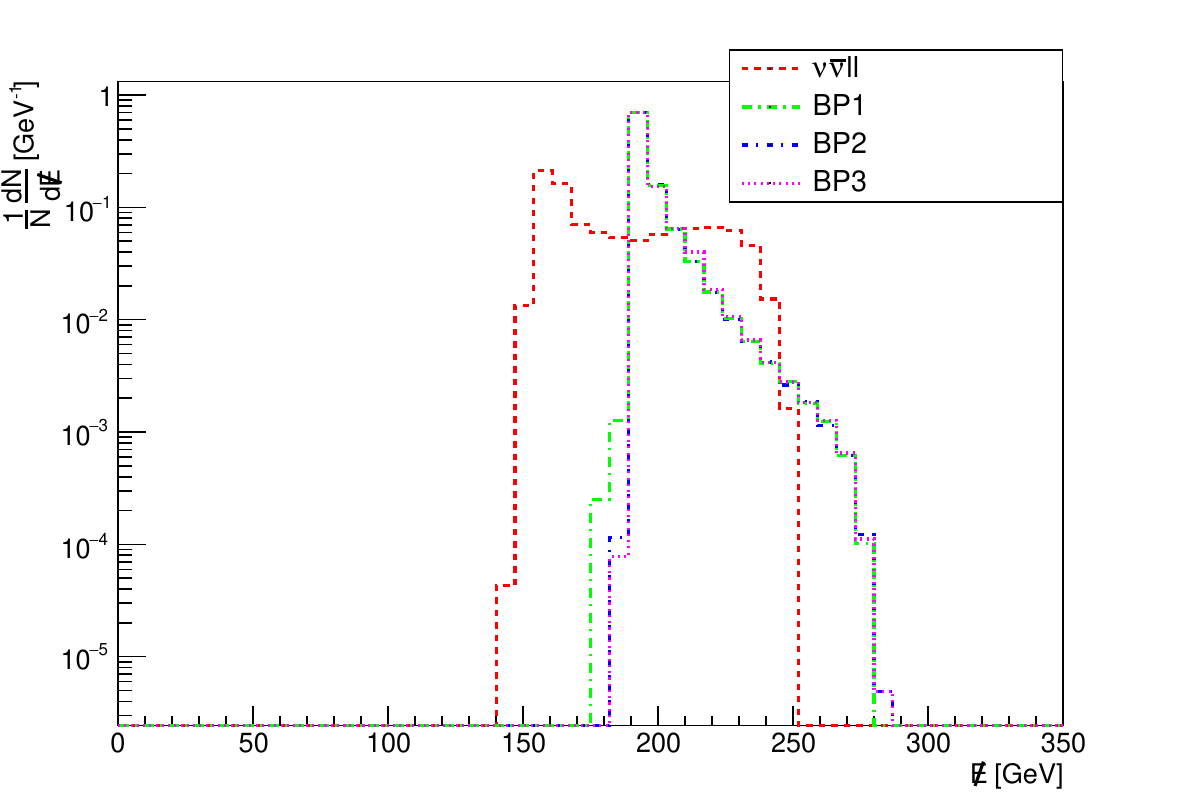}}
		\caption{Simulated signal and background events for leptonic mono-Z production at $\sqrt{s}$=240 GeV (left) and $\sqrt{s}$=365 GeV (right).		}
		\label{fig:disLep}
	\end{figure}

	The distributions of the variables which are used to select the events are shown in Fig.~\ref{fig:disLep} for both center of mass energies of 240 and 365 GeV. 	
	
	To examine the discovery potential of the model, 3 benchmark points (BP's) are defined. The specifications of BP's are shown in Tab.~\ref{tab.bps240} and Tab.~\ref{tab.bps365} for both $\sqrt{s}$ = 240 GeV, and 365 GeV, respectivley.
	
	\begin{table}[]
		\centering
		\caption{Selection cuts for different BPs in the charged leptonic channel for two FCC collision energies are shown.}
		\begin{tabular}{lcc}
			\hline
			\hline
			$\sqrt{s}$&240 GeV&365 GeV\\\hline
			Baseline selection &OSSF lepton-pairs with $P_{T_\ell} > $2 GeV,  $|\eta_\ell|$ < 3.5\\\hline
			\bf BP1\\\hline
			Cut 1& OSSF lepton-pairs with $P_{T_\ell} > $10 GeV,  $|\eta_\ell|$ < 2.8\\
			Cut 2& 80 GeV $\leq \rm m_{\ell\ell}\leq$ 100 GeV \\
			Cut 3& $\slashed{E}_T >$ 10 GeV& $\slashed{E}_T >$ 15 GeV\\
			Cut 4& $\slashed{E}>$ 120 GeV& $\slashed{E} >$ 190 GeV\\\hline
			\bf BP2\\\hline
			Cut 1& OSSF lepton-pairs with $P_{T_\ell} > $10 GeV,  $|\eta_\ell|$ < 2.8\\
			Cut 2& 80 GeV $\leq \rm m_{\ell\ell}\leq$ 100 GeV \\
			Cut 3& $\slashed{E}_T >$ 10 GeV& $\slashed{E}_T >$ 15 GeV\\
			Cut 4& $\slashed{E}>$ 140 GeV& $\slashed{E} >$ 190 GeV\\\hline
			\bf BP3\\\hline
			Cut 1& OSSF lepton-pairs with $P_{T_\ell} > $10 GeV,  $|\eta_\ell|$ < 2.8\\
			Cut 2& 80 GeV $\leq \rm m_{\ell\ell}\leq$ 100 GeV \\
			Cut 3& $\slashed{E}_T >$ 10 GeV& $\slashed{E}_T >$ 15 GeV\\
			Cut 4& $\slashed{E}>$ 140 GeV& $\slashed{E} >$ 190 GeV\\\hline
			\hline
		\end{tabular}
		\label{tab.diLepcut}
	\end{table}
	
	In Tabs.~\ref{tab.cutflow240Lep} and \ref{tab.cutflow365Lep}, the number of events and efficiencies after baseline selection and final selection cuts for both background and signal for different BP's for both $\sqrt{s}=$240 GeV and 365 GeV are shown, respectively.
	\begin{table}
		\centering
		\caption{Cut-flow table for different BPs for the leptonic mono-Z signal and background for $\sqrt{s}=$ 240 GeV FCC-ee.}
		\begin{tabular}{ccccccc}
			\hline
			\hline
			\multicolumn{7}{c}{Event Numbers (Cut efficiencies)}\\\hline
			&\multicolumn{3}{c}{$\nu\bar{\nu}\ell^+\ell^-$}&\multicolumn{3}{c}{Signal}\\
			\hline
			Selection cuts&BP1&BP2&BP3&BP1&BP2&BP3\\\hline
			Baseline Selection&&$427424$&&452527&452688&452975\\
			&&(100\%)&&(100\%)&(100\%)&(100\%)\\\hline
			After final cut&25420&4590&4590&11709&48441&55503\\
			&(5.95\%)&(1.07\%)&(1.07\%)&(2.59\%)&(10.70\%)&(12.25\%)\\\hline
			\hline
		\end{tabular}
		\label{tab.cutflow240Lep}
	\end{table}
	
	\begin{table}
		\centering
		\caption{Cut-flow table for different BPs for the leptonic mono-Z signal and background for $\sqrt{s}=$ 365 GeV FCC-ee.}
		\begin{tabular}{ccccccccc}
			\hline
			\hline
			\multicolumn{7}{c}{Event Numbers (Cut efficiencies)}\\\hline
			&\multicolumn{3}{c}{$\nu\bar{\nu}\ell^+\ell^-$}&\multicolumn{3}{c}{Signal}\\
			\hline
			Selection cuts&BP1&BP2&BP3&BP1&BP2&BP3\\\hline
			Baseline Selection&&$387150$&&468512&468208&468379\\
			&&(100\%)&&(100\%)&(100\%)&(100\%)\\\hline
			After final cut&13914&13914&13914&19614&375811&405144\\
			&(3.59\%)&(3.59\%)&(3.59\%)&(4.19\%)&(80.27\%)&(86.50\%)\\\hline
			\hline
		\end{tabular}
		\label{tab.cutflow365Lep}
	\end{table}

	\begin{table}
		\centering
		\caption{Cross sections $\sigma$ (in fb) and signal significances $\bm{S}$ after all cuts in the charged leptonic channel at $\sqrt {s}= 365$ GeV are shown. The significances are computed by assuming an integrated luminosity of $1.5~ab^{-1}$. The values in the parentheses denote the significances with 5\% background systematic uncertainty. The value of $g_v$ = 0.1.}
		\begin{tabular}{ccccc}
			\hline
			\hline
			&$M_V$(GeV)&$M_{H_2}$(GeV)&$\sigma$(fb)&$\bm{S}$ \\\hline
			BP1&30&90&736.1&155.34 (24.83)\\\hline
			BP2&50&140&96.8&288.03 (61.90)\\\hline
			BP3&60&170&26.4&124.41 (18.43)\\\hline
			$\nu_\ell \bar{\nu_\ell} \ell^+ \ell^- $&&&825\\\hline
			\hline
		\end{tabular}
		\label{tab.bps365}
	\end{table}
	\subsection{Signal Significance}
	After applying all cuts, we calculate the final signal significance for our benchmark scenarios using the definition:
	\begin{equation}
		\bm{S} = \frac{S}{\sqrt{S+\Sigma_i (B_i+\epsilon^2_i B^2_i)}}
		\label{eq1}
	\end{equation}
	where $S$ and  $B_i$ are the number of signal and ($i_{th}$ channel) background events, respectively, for a given integrated luminosity, and $\epsilon_i$ is the corresponding systematic uncertainty in the background estimation. Our results for the final signal significance for the three BPs are shown in Tabs.~\ref{tab.bps240}, \ref{tab.bps365}. We show the significance for an ideal case with zero systematic uncertainty and, in parentheses, for a more conservative case with 5\% relative systematic uncertainty applied to the dominant and subdominant backgrounds, respectively. We also tabulate the cross sections of backgrounds and signals.

	Two properties of Eq.~(\ref{eq1}) help in reading the results. For $\epsilon_i=0$ it reduces to $S/\sqrt{S+B}$ and grows as $\sqrt{L_{\rm int}}$. In the systematics-dominated regime, $\epsilon B\gg\sqrt{S+B}$, it saturates at $\bm{S}\to S/(\epsilon B)$, independent of the luminosity; for instance, from the cross sections and efficiencies of the tables the hadronic BP1 at 240~GeV has $S/B\simeq8$, so $\bm{S}\simeq150$ for $\epsilon=5\%$ is close to $(S/B)/\epsilon\simeq155$, while the leptonic BP2 at 240~GeV has $S/B\simeq0.2$ and $\bm{S}\simeq4.5$, close to $(S/B)/\epsilon\simeq4.3$. Since $S/B$ is of order one or larger for several benchmark points, the Asimov expression $\bm{S}_A=\sqrt{2\left[(S+B)\ln(1+S/B)-S\right]}$ is more accurate than $S/\sqrt{S+B}$ and can be used as a cross-check~\cite{Cowan:2010js}. With the usual conventions of $\bm{S}\simeq1.64$ for a 95\% CL exclusion and $\bm{S}=5$ for a discovery, the leptonic mode with 5\% systematics is below the exclusion level for BP1 and BP3 and below discovery for BP2 at 240~GeV, but all three benchmarks exceed $18\sigma$ at 365~GeV; the hadronic mode reaches $5\sigma$ for all benchmarks at both energies (BP3 at 240~GeV is marginal, $\bm{S}=4.98$). Because the two $Z$ decay modes are statistically independent, they can be combined in quadrature, $\bm{S}_{\rm comb}=\sqrt{\bm{S}_{\ell\ell}^2+\bm{S}_{jj}^2}$; this gives $\bm{S}_{\rm comb}\simeq533,\,155,\,47$ (BP1--BP3) at 240~GeV and $\simeq2358,\,875,\,385$ at 365~GeV without systematics, and $\simeq150,\,21.5,\,5.1$ and $\simeq966,\,156,\,42$ with 5\% systematics, i.e.\ the combination is dominated by the hadronic mode.
	
	\subsection{Hadronic mode}
	In the hadronic channel, we look at the process $e^+e^-\to VVZ, Z \to jj$, where j $\equiv$ u, d, s, c, b quarks (as shown in Fig.~\ref{diag.XVV}). The dominant SM background processes for this channel are $e^+e^-\to \nu_\ell \bar{\nu_\ell} j j$, and $e^+e^-\to \nu_\ell \ell j j$. 
	\subsubsection{Cut Based Analysis}
\begin{figure}
		\centering
		\subfloat[$\slashed{E}_T$]{\includegraphics[width=0.48\linewidth]{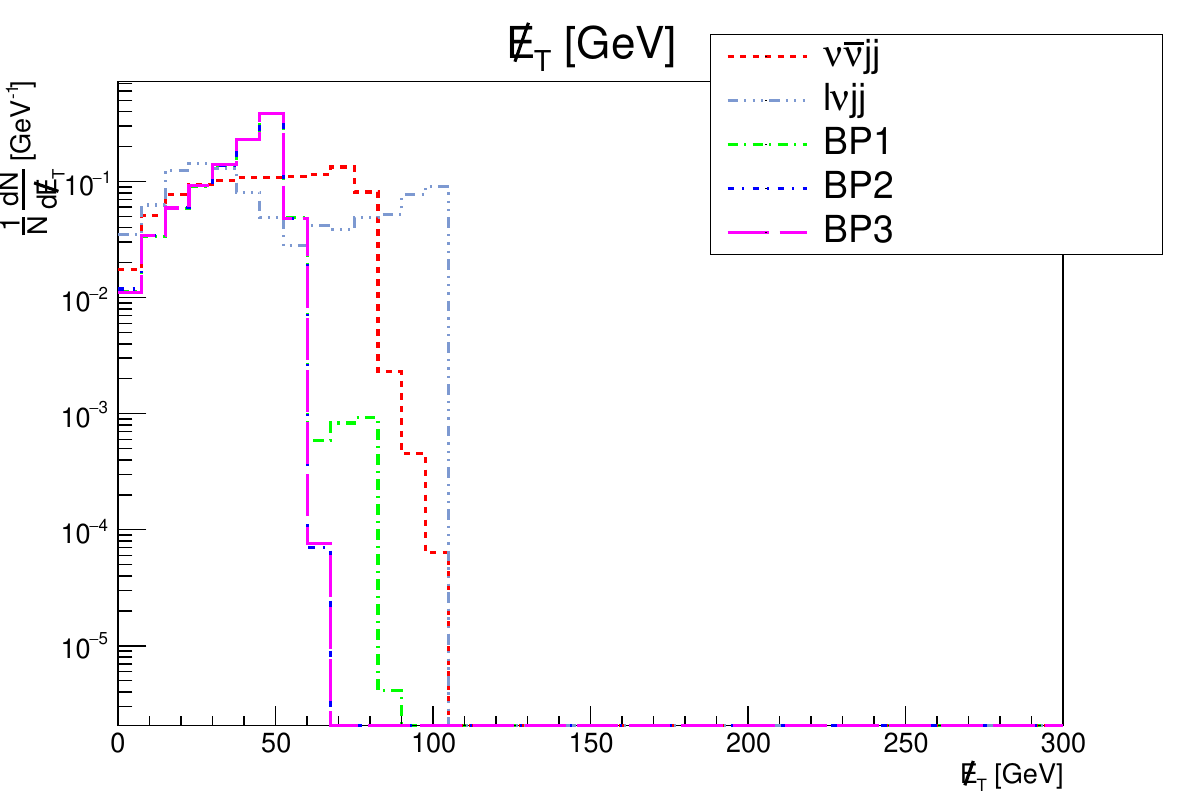}}
		\subfloat[$\slashed{E}_T$]{\includegraphics[width=0.48\linewidth]{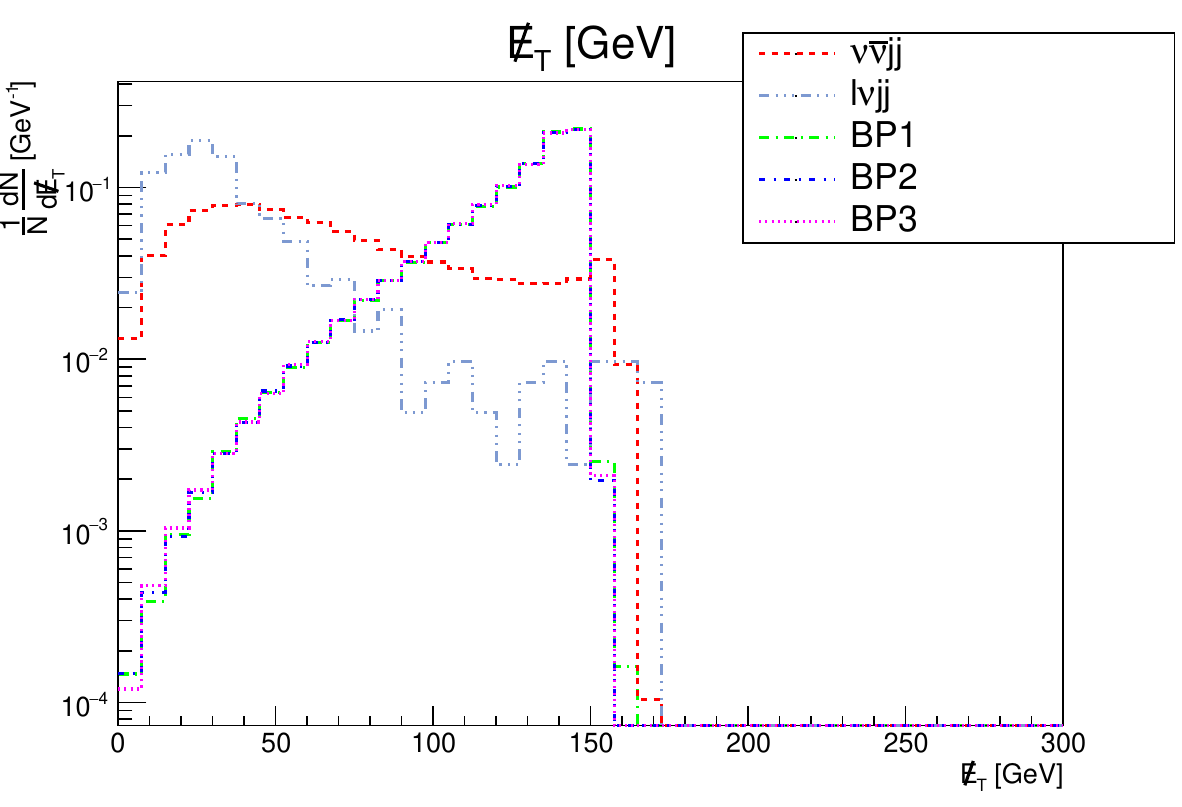}}\\
		\subfloat[M$_{jj}$]{\includegraphics[width=0.48\linewidth]{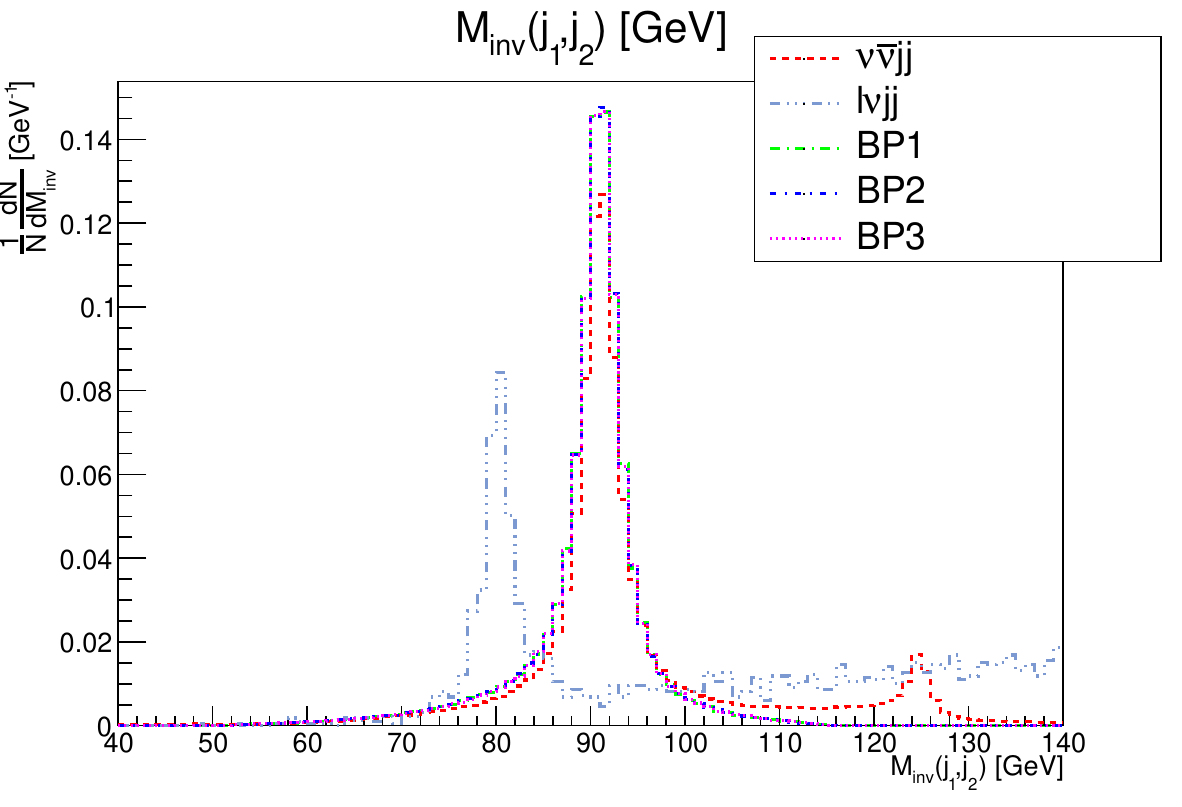}}
		\subfloat[M$_{jj}$]{\includegraphics[width=0.48\linewidth]{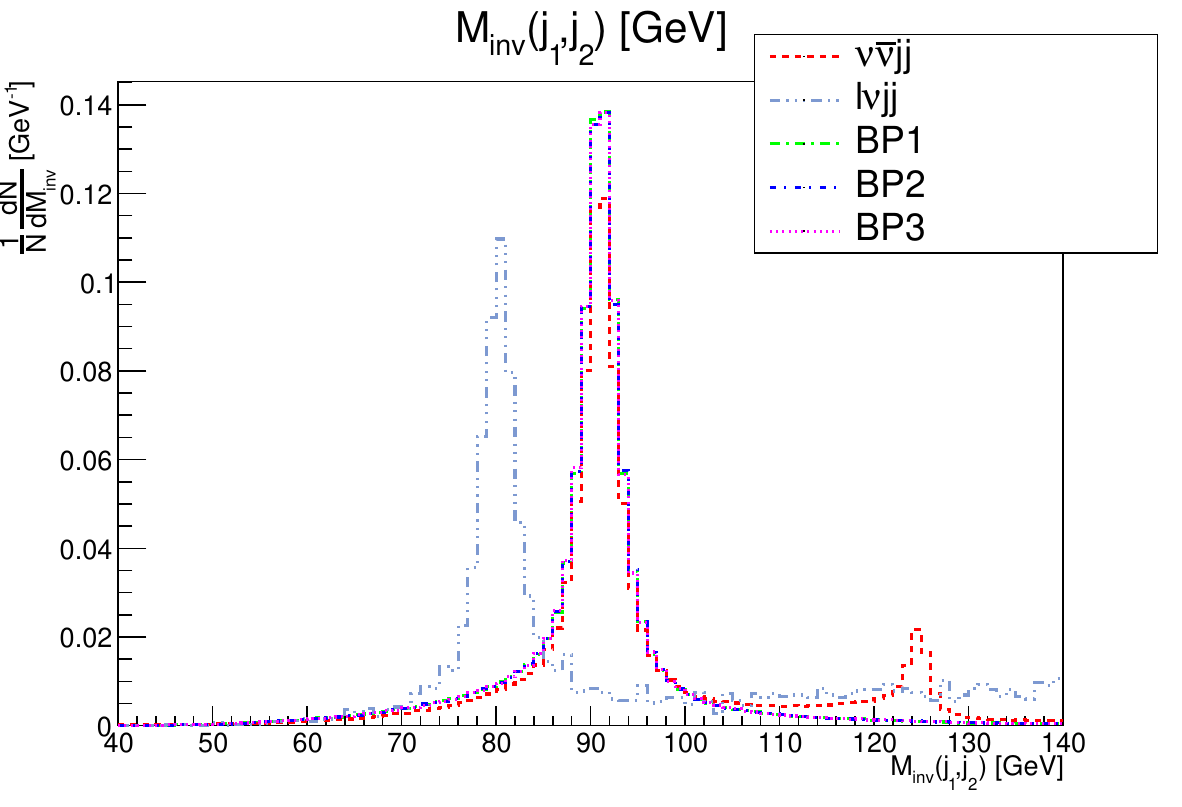}}\\
		\subfloat[MissE]{\includegraphics[width=0.48\linewidth]{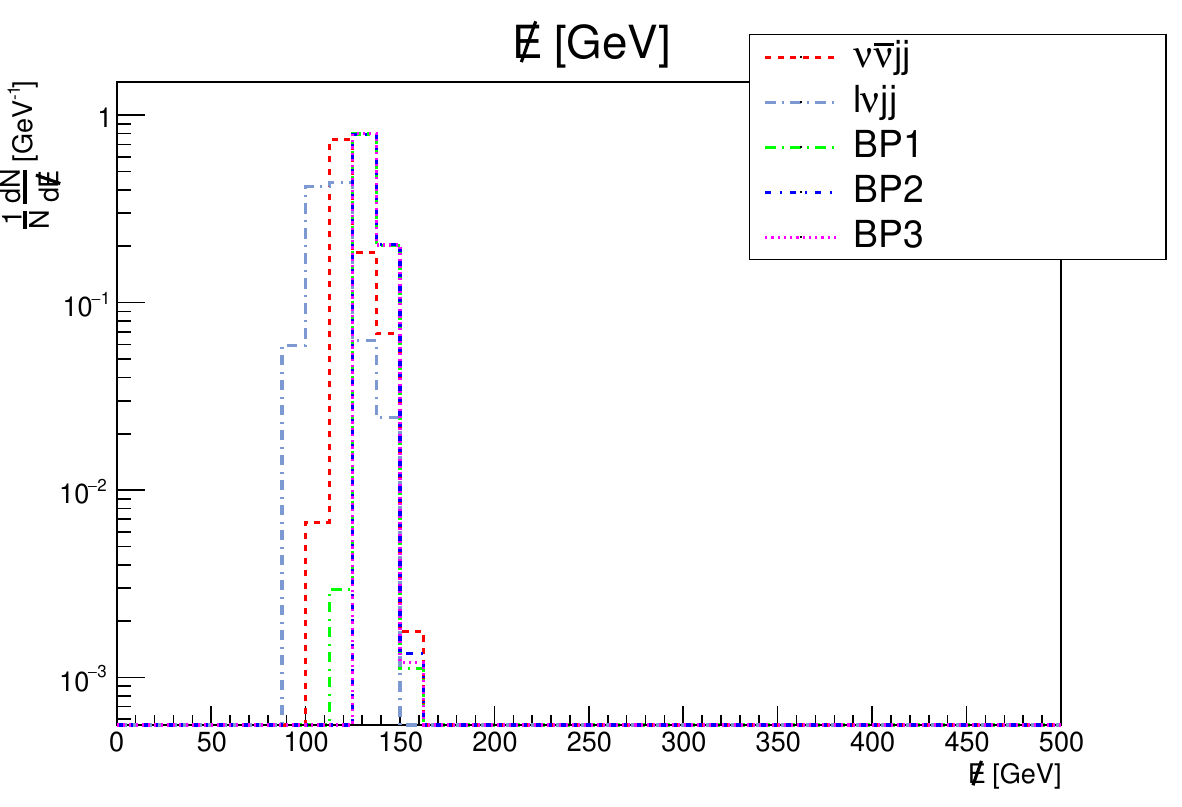}}
		\subfloat[MissE]{\includegraphics[width=0.48\linewidth]{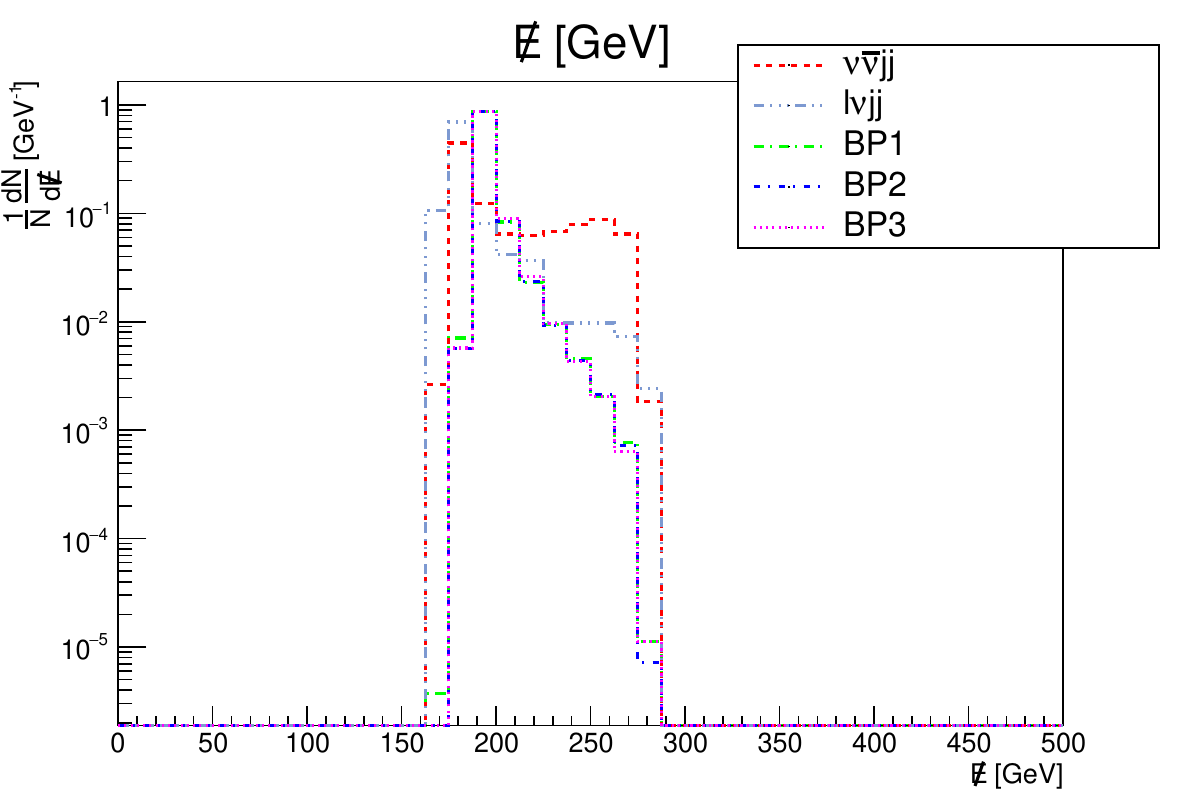}}
		\caption{Simulated signal and background events for hadronic mono-Z production at $\sqrt{s}$=240 GeV (left) and $\sqrt{s}$=365 GeV (right).}
		\label{fig:disHad}
	\end{figure}

	We simulate the events for both signals and backgrounds with basic cuts to the parameter space: $P_{T_j} > $ 5 GeV, and $|\eta_j| < $ 3.5. After obtaining cross-sections for both signal and background, as reported in Tabs.~\ref{tab.bps240Had}, \ref{tab.bps365Had} for $\sqrt{s} = $ 240 GeV and 365 GeV, respectively, we proceed with our cut-based analysis to optimize the signal significance, as we explain in more detail in the following.
	\begin{itemize}
		\item{Cut1: We select the events containing two jets with $P_{T_j} > $ 10 GeV and $|\eta_j| < $ 3.0. Veto the events if there is any other isolated lepton, photon, or jet with $P_T > $ 10 GeV and $|\eta| < $ 3.0. The two jets selected by this acceptance cut are used to reconstruct the hadronically decaying Z boson. It is realized that the $jj\nu\bar{\nu}$ is an irreducible background. Furthermore, a non-negligible amount of the $jj\ell\nu$ background survives Cut1, as the lepton $\ell$ may be clustered into a jet, have low transverse momentum, or be close to undetectable regions such as the beam direction.} 
		
		\item{Cut2: As the contribution to the $\nu\nu jj$ background comes mainly from the process $e^+ e^- \to Z + \gamma/Z$, where the di-jet comes from the $\gamma /Z$ boson, whereas the $\ell\nu jj$ background comes largely from $e^+ e^- \to W^+ W^-$, where the $W$ boson pair decays semileptonically. So, we select the events with dijet-invariant mass around the Z mass, 85 GeV $\leq M_{inv}(jj)\leq$ 100 GeV, which reduces the backgrounds considerably. To be more specific, for a $jj\ell\nu$ background with a peak around the W boson mass, it is reduced by 99\%, while 60\% of $jj\nu\bar{\nu}$ is rejected.}
		
		\item{Cut3: We select the events with $\slashed{E}_T >$ 5 GeV for $\sqrt {s}=$ 240 GeV and $\slashed{E}_T >$ 10 GeV for $\sqrt{s} =$ 365 GeV.}
		
		\item{Cut4: Same as the leptonic channel, for the $\slashed{E}$ we choose this cut to efficiently select the signal events and reject as many background events as possible.}
		The distributions of the variables that are used to select the events are shown in Fig.~\ref{fig:disHad} for both center-of-mass energies of 240 and 365 GeV.
	\end{itemize}

	Two kinematic remarks on the hadronic selection. (i)~If the dijet system with $m_{jj}\ge85$~GeV (Cut~2) is the only visible object, energy-momentum conservation implies $\slashed{E}\ge(s-m_{jj}^2)/(2\sqrt{s})\simeq105$~GeV at $\sqrt{s}=240$~GeV and $\simeq173$~GeV at $\sqrt{s}=365$~GeV, the minimum being reached for a massless recoil system. The thresholds of Tab.~\ref{tab.dijetcut} ($\slashed{E}>140$ and $190$~GeV) lie above this bound at both energies, i.e.\ they are effective, and correspond to $M_{\rm rec}\gtrsim$ 134 and 117~GeV for $m_{jj}\simeq m_Z$. (ii)~The same bound shows that the radiative-return process $e^+e^-\to q\bar q\gamma$, whose photon energy reaches $E_\gamma^{\max}\simeq103$~GeV (171~GeV) at 240 (365)~GeV, is kinematically removed by Cut~4 when the photon escapes along the beam pipe. The margin is only $\simeq19$~GeV at 365~GeV, comparable to the jet energy resolution, so since $\sigma(q\bar q\gamma)$ is very large, a cross-check with a sample generated with initial-state radiation is advisable. Finally, $t\bar t$ production is kinematically closed at 240~GeV ($\sqrt{s}<2m_t$), which explains why it is absent in Tab.~\ref{tab.cutflow240Had}, while at 365~GeV it lies only slightly above threshold.

	\begin{table}
		\centering
		\caption{Selection cuts for different BPs in the hadronic channel at two FCC collision energies are shown.}
		\begin{tabular}{lcc}
			\hline
			\hline
			$\sqrt{s}$&240 GeV&365 GeV\\\hline
			Baseline selection & $P_{T_j} > $5 GeV,  $|\eta_j|$ < 3.5, no other particles\\\hline
			\bf BP1\\\hline
			Cut 1& Exact two jets with $P_{T_j} > $10 GeV,  $|\eta_j|$ < 3.0\\
			Cut 2& 85 GeV $\leq m_{inv}(jj)\leq$ 100 GeV\\
			Cut 3& $\slashed{E}_T >$ 5 GeV& $\slashed{E}_T >$ 10 GeV\\
			Cut 4& $\slashed{E}>$ 140 GeV& $\slashed{E} >$ 190 GeV\\\hline
			\bf BP2\\\hline
			Cut 1& Exact two jets with $P_{T_j} > $10 GeV,  $|\eta_j|$ < 3.0\\
			Cut 2& 85 GeV $\leq m_{inv}(jj)\leq$ 100 GeV\\
			Cut 3& $\slashed{E}_T >$ 5 GeV& $\slashed{E}_T >$ 10 GeV\\
			Cut 4& $\slashed{E}>$ 140 GeV& $\slashed{E} >$ 190 GeV\\\hline
			\bf BP3\\\hline
			Cut 1& Exact two jets with $P_{T_j} > $10 GeV,  $|\eta_j|$ < 3.0\\
			Cut 2& 85 GeV $\leq m_{inv}(jj)\leq$ 100 GeV\\
			Cut 3& $\slashed{E}_T >$ 5 GeV& $\slashed{E}_T >$ 10 GeV\\
			Cut 4& $\slashed{E}>$ 140 GeV& $\slashed{E} >$ 190 GeV\\\hline
			\hline
		\end{tabular}
		\label{tab.dijetcut}
	\end{table}
	
	\begin{table}
		\caption{Cut-flow table for different BPs for the hadronic mono-Z signal and background for $\sqrt{s}=$ 240 GeV FCC-ee.}
		\begin{adjustbox}{width=1.1\columnwidth,center}
			\begin{tabular}{cccccccccc}
				\hline
				\hline
				\multicolumn{10}{c}{Event Numbers (Cut efficiencies)}\\\hline
				&\multicolumn{3}{c}{$\nu\bar{\nu}jj$}&\multicolumn{3}{c}{$\nu\ell jj$}&\multicolumn{3}{c}{Signal}\\
				\hline
				Selection cuts&BP1&BP2&BP3&BP1&BP2&BP3&BP1&BP2&BP3\\\hline
				Baseline Selection&&458814&&&460000&&460000&460000&420000\\
				&&(100\%)&&&(100\%)&&(100\%)&(100\%)&(100\%)\\\hline
				After final cut&&9706&&&4&&19681&20459&17967\\
				&&(2.11\%)&&&(0.0009\%)&&(4.27\%)&(4.44\%)&(4.27\%)\\\hline
				\hline
			\end{tabular}
		\end{adjustbox}
		\label{tab.cutflow240Had}
	\end{table}

	\begin{table}
		\caption{Cut-flow table for different BPs for the hadronic mono-Z signal and background for $\sqrt{s}=$ 365 GeV FCC-ee.}
		\begin{adjustbox}{width=1.1\columnwidth,center}
			\begin{tabular}{cccccccccc}
				\hline
				\hline
				\multicolumn{10}{c}{Event Numbers (Cut efficiencies)}\\\hline
				&\multicolumn{3}{c}{$\nu\bar{\nu}jj$}&\multicolumn{3}{c}{$\nu\ell jj$}&\multicolumn{3}{c}{Signal}\\
				\hline
				Selection cuts&BP1&BP2&BP3&BP1&BP2&BP3&BP1&BP2&BP3\\\hline
				Baseline Selection&&478990&&&400000&&460000&480000&460000\\
				&&(100\%)&&&(100\%)&&(100\%)&(100\%)&(100\%)\\\hline
				After final cut&&94981&&&61&&252652&264415&253740\\
				&&(19.82\%)&&&(0.015\%)&&(54.92\%)&(55.08\%)&(55.16\%)\\\hline
				\hline
			\end{tabular}
		\end{adjustbox}
		\label{tab.cutflow365Had}
	\end{table}

	\begin{table}[h!]
		\centering
		\caption{Cross sections $\sigma$ (in fb) and signal significances $\bm{S}$ after all cuts in the hadronic channel at $\sqrt {s}= 240$ GeV are shown. The significances are computed assuming an integrated luminosity of $5~ab^{-1}$. The values in the parentheses denote the significances with 5\% background systematic uncertainty. The value of $g_v$ = 0.1.}
		\begin{tabular}{ccccc}
			\hline
			\hline
			&$M_V$(GeV)&$M_{H_2}$(GeV)&$\sigma$(fb)&$\bm{S}$ \\\hline
			BP1&30&90&1620&531.47 (149.73)\\\hline
			BP2&50&140&212&149.27 (21.02)\\\hline
			BP3&60&170&57&45.22 (4.98)\\\hline
			$\nu_\ell \bar{\nu_\ell} j j$&&&421\\\hline
			$\nu_\ell \ell j j$&&&4810\\\hline
			\hline
		\end{tabular}
		\label{tab.bps240Had}
	\end{table}

	\begin{table}
		\centering
		\caption{Cross sections $\sigma$ (in fb) and signal significances $\bm{S}$ after all cuts in the hadronic channel at $\sqrt {s}= 365$ GeV are shown. The significances are computed assuming an integrated luminosity of $1.5~ab^{-1}$. The values in the parentheses denote the significances with 5\% background systematic uncertainty. The value of  $g_v$ = 0.1.}
		\begin{tabular}{ccccc}
			\hline
			\hline
			&$M_V$(GeV)&$M_{H_2}$(GeV)&$\sigma$(fb)&$\bm{S}$ \\\hline
			BP1&30&90&7445&2353.25 (965.73)\\\hline
			BP2&50&140&979&826.06 (143.52)\\\hline
			BP3&60&170&267&364.77 (37.94)\\\hline
			$\nu_\ell \bar{\nu_\ell} j j$&&&374\\\hline
			$\nu_\ell \ell j j$&&&3510\\\hline
			\hline
		\end{tabular}
		\label{tab.bps365Had}
	\end{table}

	\section{Mono-Z Channel at the ILC at $\sqrt{s}=1$ TeV}
	We study $e^+e^-\to ZVV$ at the ILC for $\sqrt{s}=1$~TeV and a total integrated luminosity of $8~\mathrm{ab}^{-1}$. We reconstruct the $Z$ boson from either a charged-lepton pair or two jets. In addition to the three benchmarks considered at the FCC-ee, the higher collision energy allows us to study BP4 and BP5, with $(M_V,M_{H_2})=(150,350)$ and $(250,600)$~GeV, respectively. We use MadGraph5 for event generation, PYTHIA8 for showering and hadronization, and Delphes3 for detector simulation by using the \textit{ILCgen} configuration card, as described in the FCC-ee analysis.

	The ILC running scenario used for the polarized samples, denoted H20, and the three beam configurations compared in this section (unpolarized, H20, and all luminosity in $(+,-)$) are defined in Sec.~\ref{subsec:polarization}.

	\subsection{Impact of Beam Polarization at the ILC}
	\label{subsec:polarization}
	
	We write the longitudinal beam polarizations in the order $(P_{e^-},P_{e^+})$. For each beam, $P=(N_R-N_L)/(N_R+N_L)$, where $N_R$ and $N_L$ are the numbers of right- and left-handed particles. We take $|P_{e^-}|=80\%$ and $|P_{e^+}|=30\%$, so that, for example, $(+,-)$ denotes a predominantly right-handed electron beam and a predominantly left-handed positron beam. If $\sigma_{h_-h_+}$ is the cross section for definite electron and positron helicities, the cross section for partially polarized beams is
	\begin{equation}
		\sigma(P_{e^-},P_{e^+}) =
		\frac{1}{4}\sum_{h_-,h_+=\pm1}
		(1+h_-P_{e^-})(1+h_+P_{e^+})\,\sigma_{h_-h_+}.
	\end{equation}
	In this expression, $P_{e^\mp}$ is given as a fraction, and $h=+1$ denotes right-handed helicity.
	
	For $e^+e^-\to W^+W^-$, the $t$-channel neutrino-exchange amplitude requires an $e^-_Le^+_R$ initial state. This helicity combination accounts for $25\%$ of unpolarized collisions, $(0.10)(0.35)=3.5\%$ for $(+,-)$, and $58.5\%$ when both polarizations are reversed. The $(+,-)$ setting therefore suppresses a major source of events with missing momentum from neutrinos. In the signal process of Fig.~\ref{diag.XVV}, a virtual $Z$ produces a visible $Z$ in association with $H_1^{(*)}$ or $H_2^{(*)}$, followed by the decay to $VV$. The initial $Zee$ vertex couples to both electron chiralities, so both opposite-helicity initial states contribute to the signal. Neutral-current backgrounds such as $ZZ$ also survive this polarization choice.
	
	The original H-20 plan covers twenty years of ILC running at $250$-$500$~GeV~\cite{Barklow:2015tja}. We use H20 for the mixed-polarization sample at the proposed $1$~TeV upgrade. We assign $40\%$ of the luminosity to each of $(+,-)$ and $(-,+)$, giving $3.2~\mathrm{ab}^{-1}$ per setting, and $10\%$ to each of $(+,+)$ and $(-,-)$, giving $0.8~\mathrm{ab}^{-1}$ per setting~\cite{Barklow:2015tja,ILC:2019gyn}. We compare H20 with two runs of the same total luminosity, one with unpolarized beams and one with all $8~\mathrm{ab}^{-1}$ in the $(+,-)$ configuration.

	In the H20 scenario, the total luminosity is split into four independent samples, which are analysed separately and combined at the level of the significance (see the signal-significance subsection below):
	\begin{center}
		\begin{tabular}{lcc}
			\hline
			$(P_{e^-},P_{e^+})$ & fraction of $L_{\rm int}$ & $L_k$ ($\mathrm{ab}^{-1}$) \\ \hline
			$(+,-)$, $(-,+)$ & 40\% each & 3.2 each \\
			$(+,+)$, $(-,-)$ & 10\% each & 0.8 each \\ \hline
		\end{tabular}
	\end{center}
	The original H-20 scenario of Ref.~\cite{Barklow:2015tja} ends at 500~GeV, so the sharing at 1~TeV is an extrapolation to the proposed energy upgrade, with the same helicity fractions as at 500~GeV. The two same-sign settings $(+,+)$ and $(-,-)$ have a small share of the luminosity; besides contributing statistically, they are included in such programmes mainly to calibrate the beam polarizations and to control systematic effects.
	
	It is important to note where the gain of H20 comes from. Merging the four samples with these weights would give almost no improvement: for the leptonic BP1, the luminosity-weighted cross sections of Tab.~\ref{tab:cross_sectionsILClep} are $\simeq9970$~fb for the signal and $\simeq825$~fb for the $\nu\bar\nu\ell^+\ell^-$ background, i.e.\ $S/B\simeq12.1$, essentially the unpolarized value of $11.6$. The benefit appears when the samples are kept separate, since their signal-to-background ratios are very different: $S/B\simeq64$ for $(+,-)$, $24$ for $(+,+)$, $6.2$ for $(-,-)$ and $5.7$ for $(-,+)$. The $(+,-)$ sample, which has the largest $S/B$ and $40\%$ of the luminosity, dominates the combined significance, and this is why the H20 values in Tabs.~\ref{tab:significance_ILC} and~\ref{tab:ILC_hadronic_significance} are close to, but below, those of the full-luminosity $(+,-)$ run when only statistical uncertainties are considered.

	The effect on the two decay modes can be seen directly in Tabs.~\ref{tab:cross_sectionsILClep} and~\ref{tab:cross_sectionsILChad}. Before the analysis cuts, changing from unpolarized beams to $(+,-)$ reduces the $\nu\bar{\nu}\ell^+\ell^-$ cross section from $752.4$ to $199$~fb. In the hadronic mode the $\nu\bar{\nu}jj$ rate falls from $810$ to $140$~fb and the $\ell\nu jj$ rate from $1340$ to $240$~fb. Using the BP1 signal cross sections in the same tables, the ratio of signal to total background increases by a factor of $5.5$ in the leptonic mode and $8.3$ in the hadronic mode.
	
	At this energy $WW$ and $ZZ$ fusion also give access to $VV$ production through the complementary final states $\nu\bar\nu VV$ and $e^+e^-VV$, respectively~\cite{Buttazzo:2018qqp,Costantini:2020stv,Ruiz:2021tdt}.

	\subsection{Leptonic mode}
	
	For $Z\to e^+e^-$ or $\mu^+\mu^-$, the two charged leptons identify the visible $Z$ and the $VV$ pair carries away the recoil momentum. We include the SM background $e^+e^-\to\ell^+\ell^-\nu\bar{\nu}$. It contains $ZZ$ events with one invisible $Z$ decay, as well as $W^+W^-$ events in which the two charged leptons come from different $W$ bosons. The former have a dilepton mass peak in the signal region, while the latter populate a broader range of $m_{\ell\ell}$.
	
	\subsubsection{Cut Based Analysis}
	
	The baseline sample contains an opposite-sign, same-flavor (OSSF) lepton pair with $p_{T,\ell}>2$~GeV and $|\eta_\ell|<3.5$. We apply the following four cuts to all five benchmarks, as summarized in Tab.~\ref{tab:cutflowILClep}.
	\begin{itemize}
		\item Cut~1 selects exactly two charged leptons forming one OSSF pair. Each lepton must satisfy $p_{T,\ell}>10$~GeV and $|\eta_\ell|<2.8$.
		\item Cut~2 requires $80~\mathrm{GeV}\leq m_{\ell\ell}\leq100~\mathrm{GeV}$. This window retains the $Z$ peak and rejects part of the $W^+W^-$ continuum.
		\item Cut~3 requires $\slashed{E}_T=\big|\sum_{\rm vis}\vec{p}_T\big|>20$~GeV. It suppresses events in which the missing momentum points close to the beam axis.
		\item Cut~4 requires $\slashed{E}=\sqrt{s}-\sum_{\rm vis}E_i>500$~GeV. Here the sum runs over the reconstructed visible objects, so that this cut tests the energy carried by the recoil system.
	\end{itemize}
	
	\begin{figure}
		\centering
		\subfloat[M$_{ll}$]{\includegraphics[width=0.48\linewidth]{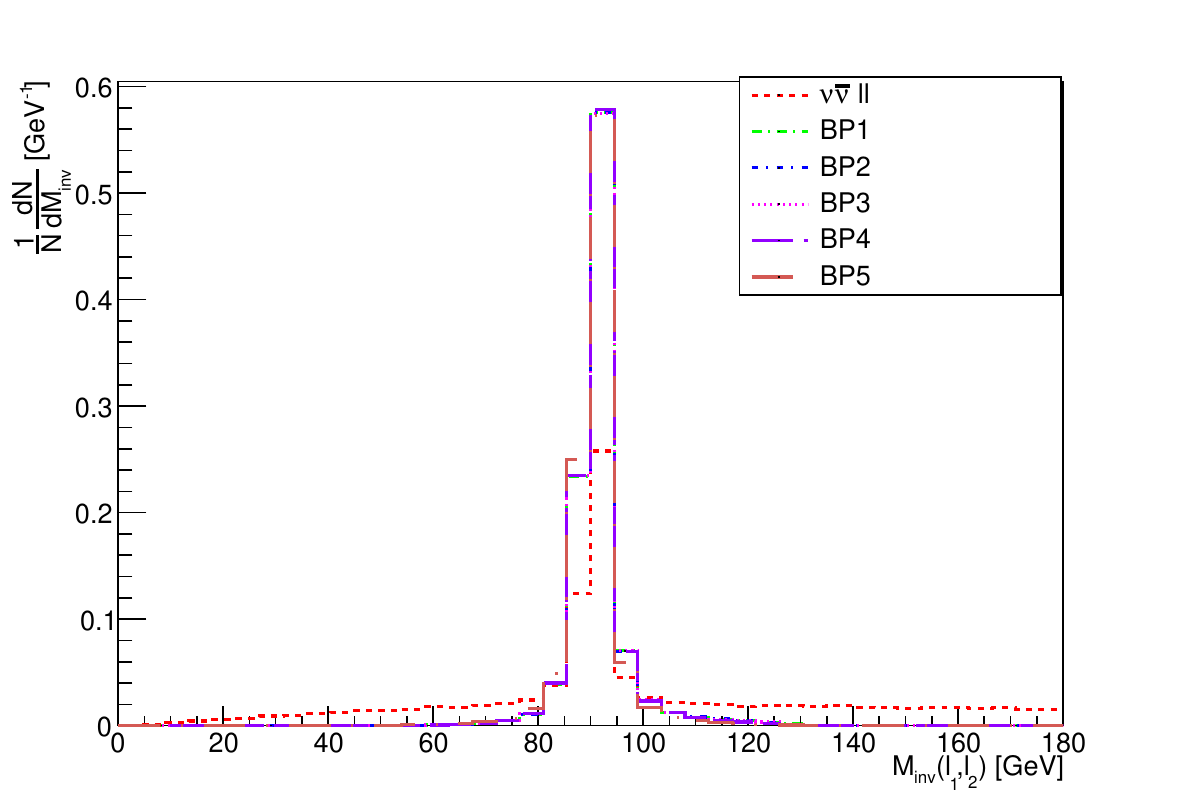}}
		\subfloat[$\slashed{E}_T$]{\includegraphics[width=0.48\linewidth]{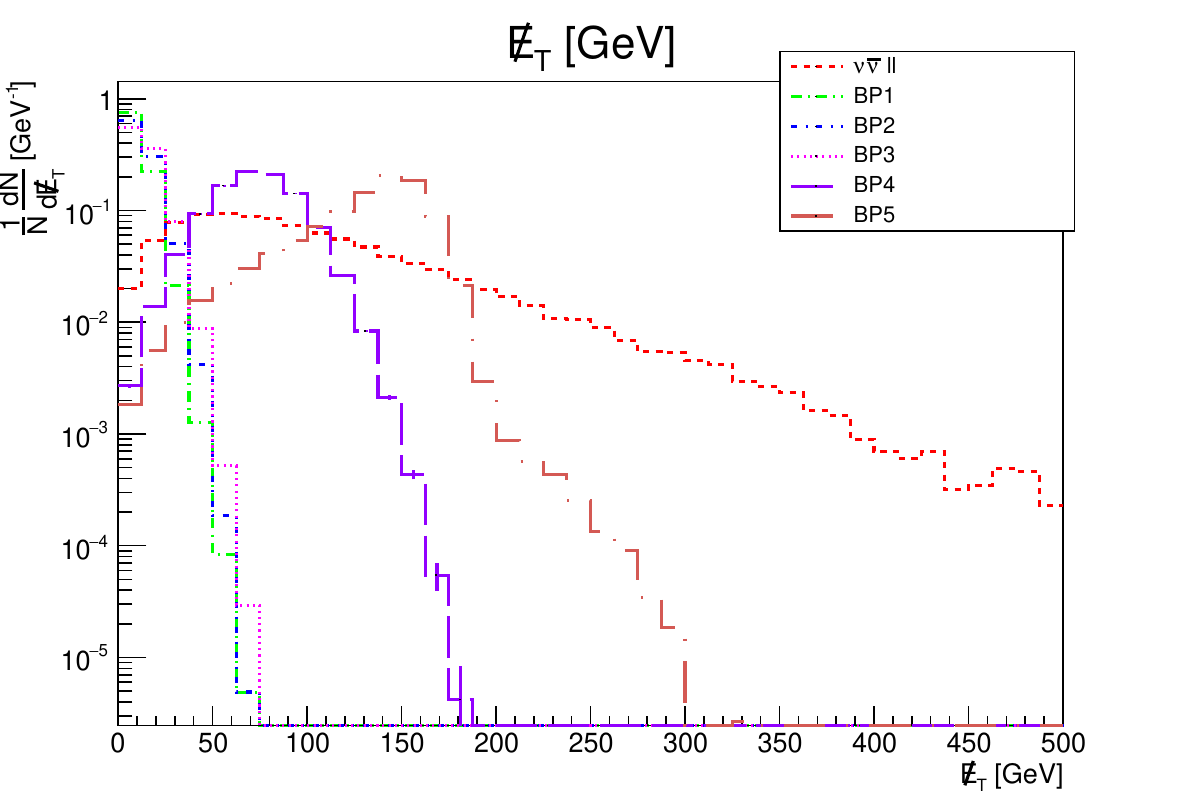}}\\
		\subfloat[$\slashed{E}$]{\includegraphics[width=0.48\linewidth]{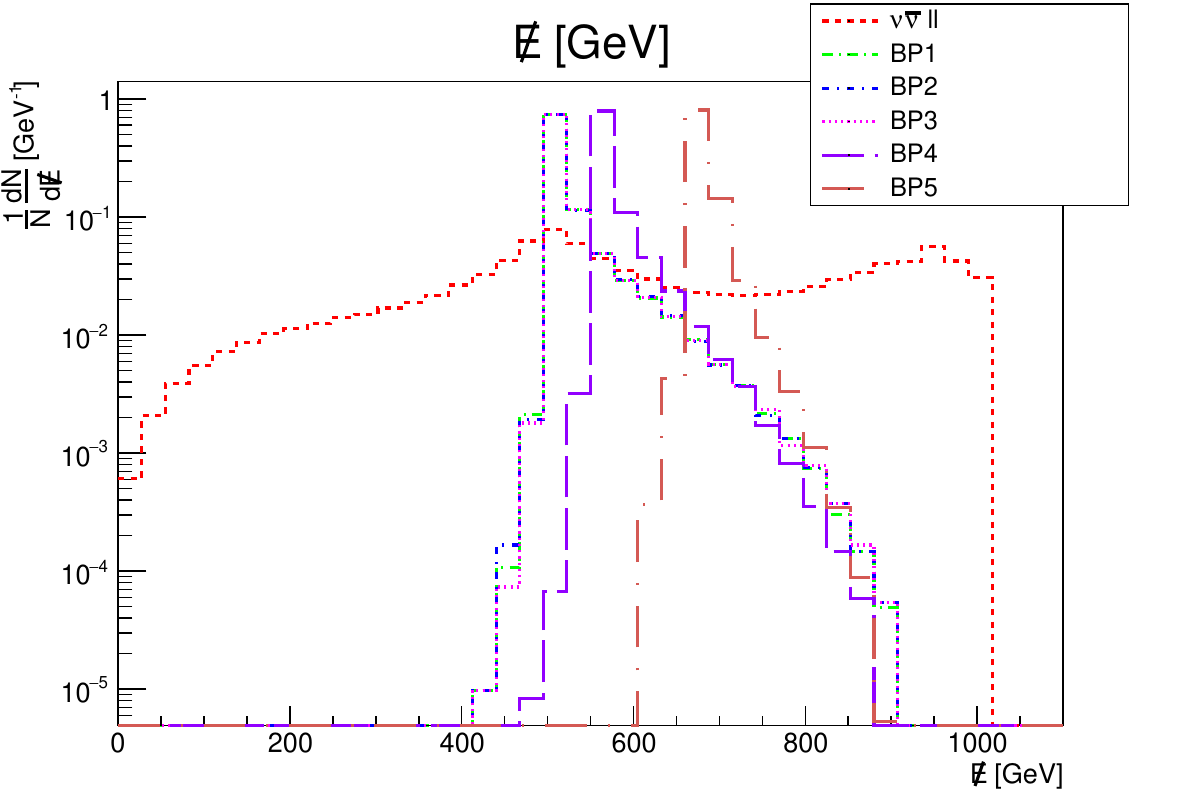}}
		\caption{Simulated signal and background events for leptonic mono-Z production (unpolarized) at $\sqrt{s}$= 1 TeV ILC.}
		\label{fig:disLepILC}
	\end{figure}
	
	Figure~\ref{fig:disLepILC} shows the three distributions used after lepton selection. Of the $273\,988$ background events in the baseline sample, $32\,165$ survive all four cuts ($11.74\%$). The corresponding signal efficiencies are $5.76\%$, $11.55\%$, and $17.12\%$ for BP1--BP3, and rise to $92.62\%$ and $93.73\%$ for BP4 and BP5 (Tab.~\ref{tab:ILC_leptonic_cutflow}). These percentages are relative to the baseline selection.
	
	\begin{table}[b]
		\centering
		\caption{Cut-flow table for different BPs for the leptonic mono-Z signal and background for $\sqrt{s}=1$ TeV ILC.}
		\label{tab:ILC_leptonic_cutflow}
		\resizebox{\textwidth}{!}{
			\begin{tabular}{lcccccc}
				\hline\hline
				& \multicolumn{1}{c}{SM Background}
				& \multicolumn{5}{c}{Signal} \\
				\cline{2-2} \cline{3-7}
				Selection cuts
				& $\nu\bar{\nu}\ell^+\ell^-$
				& BP1 & BP2 & BP3 & BP4 & BP5 \\
				\hline
				Baseline Selection
				& 273988
				& 219302 & 219358 & 219598 & 255383 & 399699 \\
				& (100\%)
				& (100\%) & (100\%) & (100\%) & (100\%) & (100\%) \\
				
				After final cut
				& 32165
				& 12626 & 25337 & 37585 & 236554 & 374624 \\
				& (11.74\%)
				& (5.76\%) & (11.55\%) & (17.12\%)
				& (92.62\%) & (93.73\%) \\
				\hline\hline
			\end{tabular}
		}
	\end{table}
	
	\begin{table}[t!]
		\centering
		\caption{Selection cuts for different BPs in the charged leptonic channel at $\sqrt{s}$ = 1 TeV ILC  are shown.}
		\begin{adjustbox}{width=\columnwidth,center}
			\begin{tabular}{lccccc}
				\hline
				\hline
				$\sqrt{s}$&&1 TeV\\\hline
				Baseline selection&&OSSF lepton-pairs with $P_{T_\ell} >$ 2 GeV, $|\eta_\ell| < $ 3.5\\\hline
				Selection cuts & BP1 & BP2 & BP3 & BP4 & BP5 \\
				\hline
				Cut 1              &&Exactly 2 OSSF leptons with $P_{T_\ell} >$ 10 GeV, $|\eta_\ell| < $ 2.8&&&  \\
				Cut 2           &  & 80 GeV $\leq \rm m_{\ell\ell}\leq$ 100 GeV &  & &  \\
				Cut 3                  &   &  $\slashed{E}_T > $ 20 GeV &       &     &        \\
				Cut 4                  &     &   $\slashed{E} > $ 500 GeV   &    &    &        \\
				\hline
				\hline
			\end{tabular}
		\end{adjustbox}
		\label{tab:cutflowILClep}
	\end{table}

	The position of Cut~4 helps explain the difference between the light and heavy benchmarks. When the lepton pair accounts for the entire visible system, $M_{\rm rec}^2=2\sqrt{s}\,\slashed{E}-s+m_{\ell\ell}^2$. At $1$~TeV the requirement $\slashed{E}>500$~GeV therefore selects $M_{\rm rec}>m_{\ell\ell}$. The $90$~GeV $H_2$ resonance of BP1 lies close to this boundary for an on-shell $Z$, whereas the minimum $VV$ invariant masses of BP4 and BP5 are $300$ and $500$~GeV. Initial-state radiation, beamstrahlung, and finite energy resolution can shift the reconstructed recoil mass away from the true $VV$ invariant mass.

	\begin{table}[t]
		\centering
		\caption{Comparison of the leptonic mono-Z background and signal cross sections for different choices of beam polarization at $\sqrt{s}$ = 1 TeV ILC.}
		\begin{adjustbox}{width=1.1\columnwidth,center}
			\begin{tabular}{cccccccccc}
				\toprule
				&&&&\multicolumn{5}{r}{Polarized cross section (fb)}\\
				\cmidrule(lr){6-9}
				Benchmarks &
				$M_V$ (GeV) &
				$M_{H_2}$ (GeV) &
				Unpolarized cross section (fb) &
				Polarization P$(e^-,e^+)$&
				$(+,+)$ &
				$(+,-)$ &
				$(-,+)$ &
				$(-,-)$ &\\
				\midrule
				BP1&30&90&8720&(80,30)&7500&12740&8870&5750\\\hline
				BP2&50&140&1140&(80,30)&988&1670&1170&758\\\hline
				BP3&60&170&317&(80,30)&273&464&324&209\\\hline
				BP4&150&350&0.79&(80,30)&0.68&1.16&0.81&0.52\\\hline
				BP5&250&600&0.056&(80,30)&0.048&0.082&0.057&0.037\\
				\midrule
				$\nu\bar{\nu}\ell^+\ell^-$ &&&752.4&(80,30)&318&199&1550&934\\
				\bottomrule
			\end{tabular}
		\end{adjustbox}
		\label{tab:cross_sectionsILClep}
	\end{table}
	
	\begin{table}[t!]
		\centering
		\caption{Signal significance in the leptonic mono-$Z$ channel at
			$\sqrt{s}=1~\mathrm{TeV}$ with an integrated luminosity of
			$L_{\mathrm{int}}=8~\mathrm{ab}^{-1}$. The values in parentheses
			denote the significance including the background systematic
			uncertainty.}
		\begin{adjustbox}{width=1.1\columnwidth,center}
			\begin{tabular}{cccccccccccccccc}
				\hline
				& \multicolumn{5}{c}{Unpolarized beams}
				& \multicolumn{5}{c}{H20 scenario}
				& \multicolumn{5}{c}{Optimally polarized beams (+,-)} \\
				\cline{2-16}
				& BP1 & BP2 & BP3 & BP4 & BP5
				& BP1 & BP2 & BP3 & BP4 & BP5
				& BP1 & BP2 & BP3 & BP4 & BP5 \\
				\hline
				& 1201.76& 501.41 & 250.71 & 4.78 & 0.54
				& 1300 & 597 & 339 & 10.7 & 1.23
				& 243.04 & 148.45 & 80.21 & 0.38 & 0.047 \\
				& (90.62) & (23.82) & (9.83) & (0.15) & (0.02)
				& (630) & (194) & (82.3) & (1.33) & (0.15)
				& (33.33) & (16.98) & (7.99) & (0.032) & (0.003) \\
				\hline
			\end{tabular}
		\end{adjustbox}
		\label{tab:significance_ILC}
	\end{table}
	
	\subsection{Hadronic mode}
	
	For $Z\to q\bar q$, with $q=u,d,s,c,b$, we reconstruct the $Z$ candidate from two jets. We include $e^+e^-\to\nu\bar{\nu}jj$, which has the same visible final state as the signal, and $e^+e^-\to\ell\nu jj$, including the charge-conjugate mode. The latter contributes when the charged lepton escapes acceptance or fails identification. Top-pair production, $e^+e^-\to t\bar{t}\to b\bar b W^+W^-$, typically gives additional jets or charged leptons. We therefore treat $t\bar{t}$ as a minor background to the exclusive mono-$Z$ selection and omit it from the ILC cut flows and significances.
	
	\begin{table}[t]
		\centering
		\caption{Comparison of the hadronic mono-$Z$ background and signal cross sections for different choices of beam polarization at $\sqrt{s}$ = 1 TeV ILC.}
		\begin{adjustbox}{width=1.1\columnwidth,center}
			\begin{tabular}{cccccccccc}
				\toprule
				&&&&\multicolumn{5}{r}{Polarized cross section (fb)}\\
				\cmidrule(lr){6-9}
				Benchmarks &
				$M_V$ (GeV) &
				$M_{H_2}$ (GeV) &
				Unpolarized cross section (fb) &
				Polarization P$(e^-,e^+)$&
				$(+,+)$ &
				$(+,-)$ &
				$(-,+)$ &
				$(-,-)$ &\\
				\midrule
				BP1&30&90&91750&(80,30)&78880&134000&93480&60550\\\hline
				BP2&50&140&12100&(80,30)&10410&17650&12330&7900\\\hline
				BP3&60&170&3340&(80,30)&2880&4880&3400&2200\\\hline
				BP4&150&350&8&(80,30)&6.9&11.7&8.2&5.3\\\hline
				BP5&250&600&0.5&(80,30)&0.48&0.82&0.57&0.37\\
				\midrule
				$\nu\bar{\nu}jj$ &&&810&(80,30)&220&140&1860&1000\\\hline
				$\nu_\ell\ell jj$ &&&1340&(80,30)&430&240&2990&1680\\
				\bottomrule
			\end{tabular}
		\end{adjustbox}
		\label{tab:cross_sectionsILChad}
	\end{table}
	
	\subsubsection{Cut Based Analysis}
	
	At baseline, we require jets with $p_{T,j}>5$~GeV and $|\eta_j|<3.5$, with no additional identified visible object. The selection in Tab.~\ref{tab:cutflowILChad} uses the same missing-energy thresholds as the leptonic analysis.
	\begin{itemize}
		\item Cut~1 requires exactly two jets with $p_{T,j}>20$~GeV and $|\eta_j|<3.0$. We add their four-momenta to reconstruct the dijet candidate.
		\item Cut~2 selects $85~\mathrm{GeV}\leq m_{jj}\leq100~\mathrm{GeV}$. The lower edge lies above $m_W$, reducing the contribution from semileptonic $W^+W^-$ decays in $\ell\nu jj$. Events with a genuine $Z\to jj$ in $\nu\bar\nu jj$ populate the same mass window as the signal.
		\item Cut~3 requires $\slashed{E}_T>20$~GeV. When the dijet accounts for all visible momentum, this is also a lower bound on the transverse momentum of the dijet system.
		\item Cut~4 retains events with $\slashed{E}>500$~GeV. As in the leptonic mode, the energy is calculated from all reconstructed visible objects. For events in which only the dijet is visible, $m_{jj}$ replaces $m_{\ell\ell}$ in the recoil-mass relation.
	\end{itemize}
	
	\begin{table}
		\centering
		\caption{Signal significance in the hadronic mono-$Z$
			channel at $\sqrt{s}=1$ TeV with an integrated luminosity
			of $\mathcal{L}_{\mathrm{int}}=8$ ab$^{-1}$. The values
			in parentheses denote the significance including the
			5\% background systematic uncertainty.}
		\label{tab:ILC_hadronic_significance}
		\resizebox{\textwidth}{!}{
			\begin{tabular}{ccccccccccccccc}
				\hline\hline
				\multicolumn{5}{c}{Unpolarized beams}
				& \multicolumn{5}{c}{H20 scenario}
				& \multicolumn{5}{c}{Optimally polarized beams (+,-)} \\
				\cline{1-5} \cline{6-10} \cline{11-15}
				BP1 & BP2 & BP3 & BP4 & BP5
				& BP1 & BP2 & BP3 & BP4 & BP5
				& BP1 & BP2 & BP3 & BP4 & BP5 \\
				\hline
				3121.18 & 762.40 & 247.39 & 0.743 & 0.221
				& 3377.15 & 990.17 & 412.89 & 2.644 & 0.586
				& 4104.22 & 1358.49 & 601.05 & 3.994 & 0.885 \\
				
				(105.08) & (13.76) & (3.82) & (0.011) & (0.003)
				& (923.10) & (124.86) & (34.89) & (0.150) & (0.033)
				& (893.55) & (116.85) & (32.73) & (0.141) & (0.031) \\
				\hline\hline
			\end{tabular}
		}
	\end{table}
	
	\begin{table}[h]
		\centering
		\caption{Cut-flow table for different BPs for the hadronic mono-Z signal and background for $\sqrt{s}=$ 1 TeV ILC.}
		\label{tab:ILC_hadronic_cutflow}
		\resizebox{\textwidth}{!}{
			\begin{tabular}{lccccccc}
				\hline\hline
				& \multicolumn{2}{c}{SM Backgrounds}
				& \multicolumn{5}{c}{Signal} \\
				\cline{2-3} \cline{4-8}
				Selection cuts
				& $\nu\bar{\nu}jj$
				& $\ell\nu jj$
				& BP1 & BP2 & BP3 & BP4 & BP5 \\
				\hline
				Baseline Selection
				& 499985 & 500000
				& 500000 & 500000 & 500000 & 334728 & 500000 \\
				
				& (100\%) & (100\%)
				& (100\%) & (100\%) & (100\%) & (100\%) & (100\%) \\
				
				After final cut
				& 171909 & 17545
				& 8091 & 8031 & 8065 & 9372 & 44559 \\
				
				& (34.38\%) & (3.51\%)
				& (1.62\%) & (1.61\%) & (1.61\%) & (2.80\%) & (8.91\%) \\
				\hline\hline
			\end{tabular}
		}
	\end{table}
	
\begin{figure}
		\centering
		\subfloat[M$_{jj}$]{\includegraphics[width=0.48\linewidth]{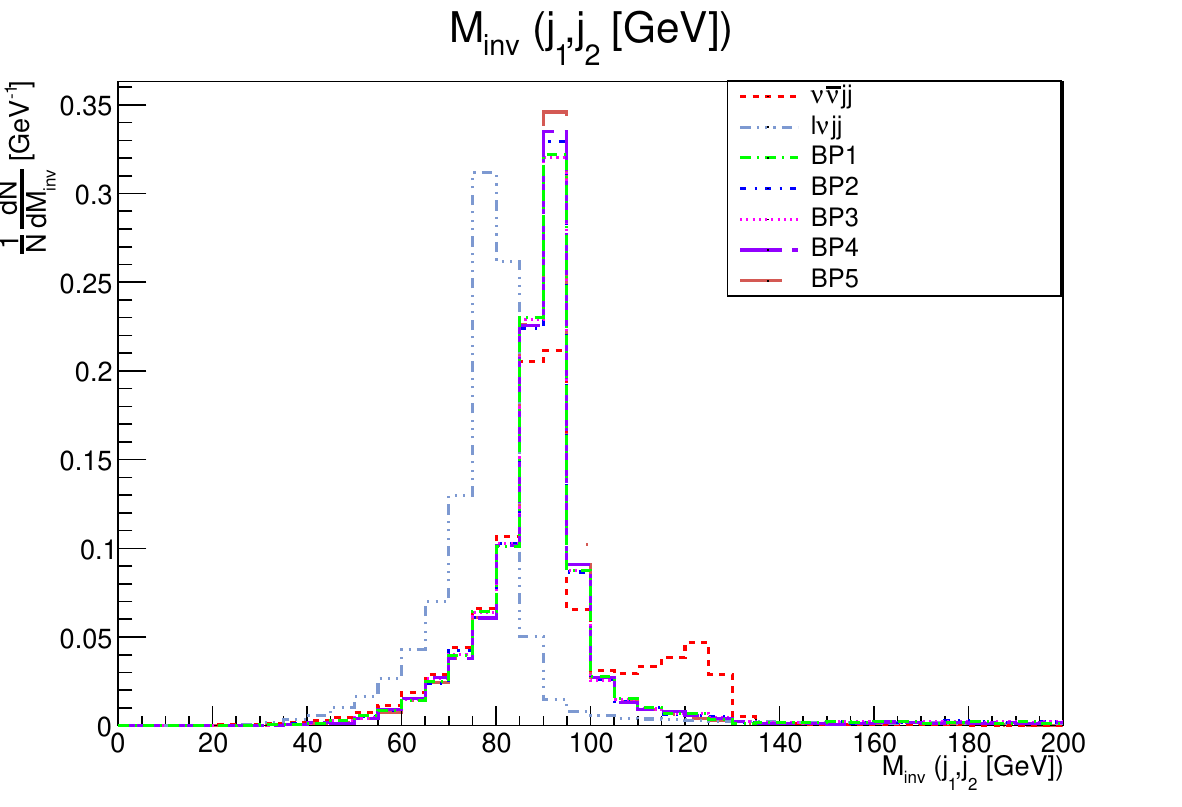}}
		\subfloat[$\slashed{E}_T$]{\includegraphics[width=0.48\linewidth]{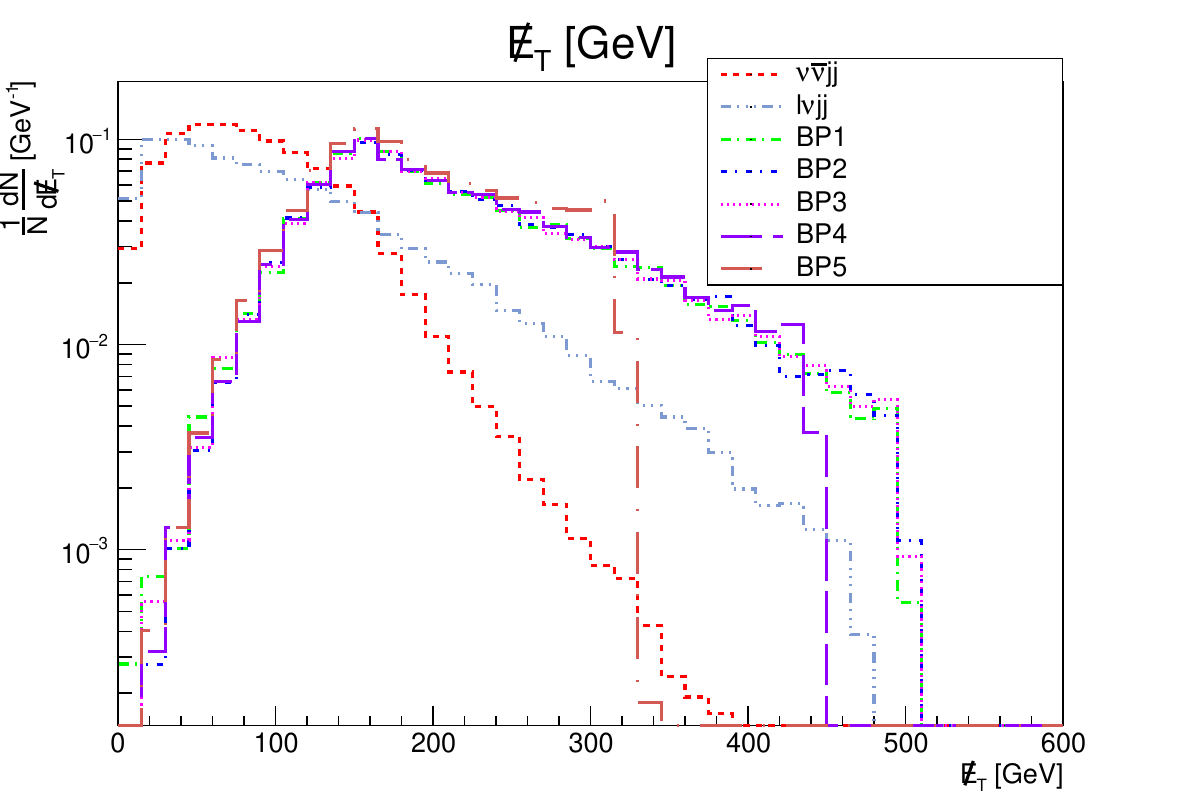}}\\
		\subfloat[$\slashed{E}$]{\includegraphics[width=0.48\linewidth]{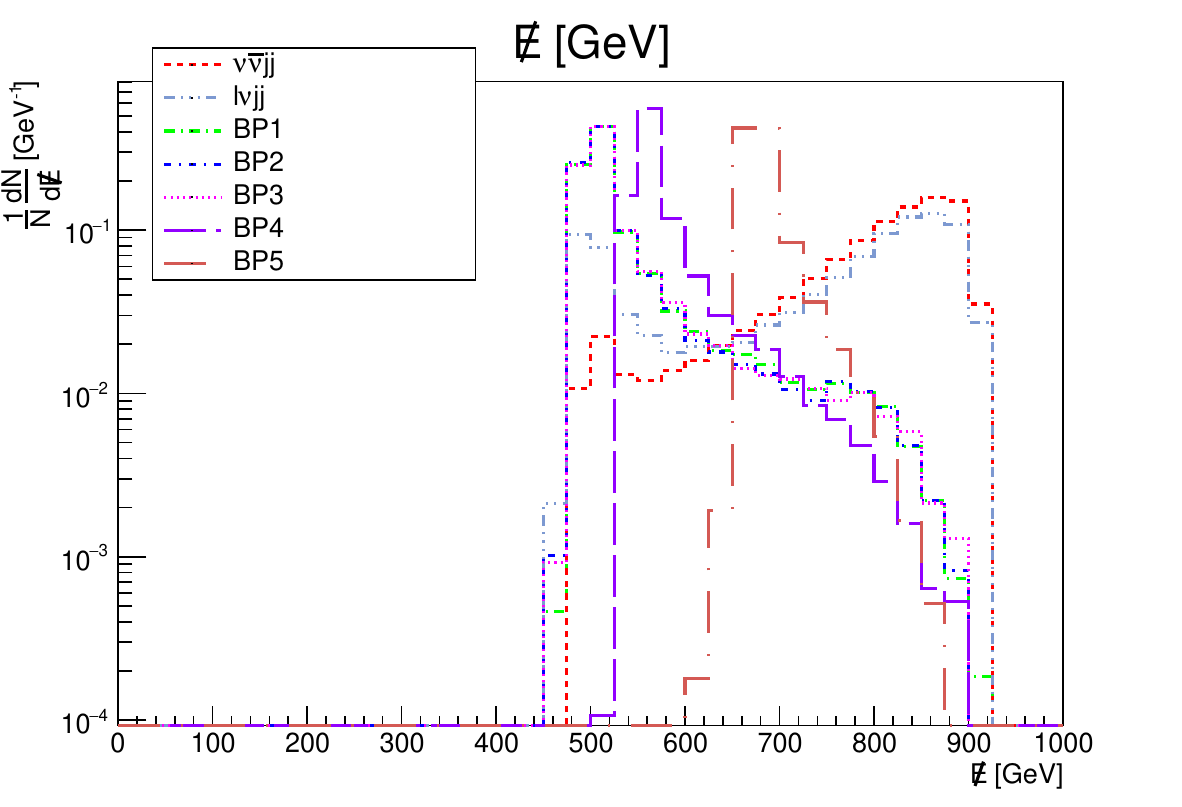}}
		\caption{Simulated signal and background events for hadronic mono-Z production (unpolarized) at $\sqrt{s}$= 1 TeV ILC.}
		\label{fig:disHadILC}
		\label{fig:ILC_unpol_dijet}
	\end{figure}

	The full selection retains $34.38\%$ of the baseline $\nu\bar\nu jj$ sample and $3.51\%$ of the $\ell\nu jj$ sample (Tab.~\ref{tab:ILC_hadronic_cutflow}). Although $\ell\nu jj$ has the larger unpolarized cross section before selection, $\nu\bar\nu jj$ supplies most of the final hadronic background. The signal efficiencies relative to baseline are about $1.6\%$ for BP1--BP3, $2.80\%$ for BP4, and $8.91\%$ for BP5. Figure~\ref{fig:ILC_unpol_dijet} shows the dijet mass, missing transverse energy, and missing energy used in this selection.
	
	\begin{table}[t!]
		\centering
		\caption{Selection cuts for different BPs in the hadronic channel at $\sqrt{s}$ = 1 TeV ILC  are shown.}
		\begin{adjustbox}{width=\columnwidth,center}
			\begin{tabular}{lccccc}
				\hline
				\hline
				$\sqrt{s}$&&1 TeV\\\hline
				Baseline selection&& $P_{T_j} >$ 5 GeV, $|\eta_j| < $ 3.5, no other particles\\\hline
				Selection cuts & BP1 & BP2 & BP3 & BP4 & BP5 \\
				\hline
				Cut 1              &&Exactly 2-jets with $P_{T_j} >$ 20 GeV, $|\eta_j| < $ 3.0&&&  \\
				Cut 2           &  & 85 GeV $\leq \rm m_{jj}\leq$ 100 GeV &  & &  \\
				Cut 3                  &   &  $\slashed{E}_T > $ 20 GeV &       &     &        \\
				Cut 4                  &     &   $\slashed{E} > $ 500 GeV   &    &    &        \\
				\hline
				\hline
			\end{tabular}
		\end{adjustbox}
		\label{tab:cutflowILChad}
	\end{table}
	
	\subsection{Signal Significance}
	We use Eq.~(\ref{eq1}) for each setting separately. For H20, each yield uses the luminosity assigned to that setting. We then combine the four significances in quadrature~\cite{Kundu:2023},
	\begin{equation}
		\bm{S}_{H20}=\sqrt{\bm{S}^2_{(+,+)} + \bm{S}^2_{(+,-)} + \bm{S}^2_{(-,+)} + \bm{S}^2_{(-,-)}}.
	\end{equation}
	
	Tables~\ref{tab:significance_ILC} and~\ref{tab:ILC_hadronic_significance} give the results for the leptonic and hadronic modes, respectively. 	
	
	The values in parentheses include a $5\%$ normalization uncertainty for each background in Eq.~(\ref{eq1}). In the H20 combination, we treat these uncertainties as independent between beam settings. Under this assumption, BP1--BP3 retain $\bm S>5$ in both decay modes for H20. For BP4, the H20 significance falls to $1.33$ in the leptonic mode and $0.150$ in the hadronic mode. Normalization errors shared by several beam settings would require a joint fit with common nuisance parameters.

	\newpage
	\section{Conclusion}
	We have studied the mono-$Z$ signature of Higgs-portal VDM at the FCC-ee ($\sqrt{s}=240$ and $365$~GeV) and at the ILC ($\sqrt{s}=1$~TeV), with a cut-based analysis of both leptonic and hadronic $Z$ decays.
	
	For $g_v=0.1$, $M_V=30$--$60$~GeV and $M_{H_2}=90$--$170$~GeV (BP1--BP3), the hadronic mode at the FCC-ee is the more sensitive one, with $\bm{S}\simeq531$, 149, 45 at 240~GeV and $2353$, 826, 365 at 365~GeV ($150$, 21, 5.0 and $966$, 144, 38 with 5\% background systematics). The leptonic mode gives $\bm{S}\simeq36$, 43, 13 at 240~GeV ($1.5$, $4.5$, $1.2$ with systematics) and $155$, 288, 124 at 365~GeV ($24.8$, $61.9$, $18.4$). This is due to the roughly ten times larger $Z\to jj$ rate and to $S/B\simeq8$ after the selection, against $S/B\lesssim1$ in the leptonic mode. With 5\% systematics, only the hadronic channel reaches $5\sigma$ at 240~GeV, while the leptonic channel remains a cleaner, independent cross-check.
	
	At the ILC with $8~\mathrm{ab}^{-1}$, beam polarization mainly improves $S/B$, by about 5.5 (leptonic) and 8.3 (hadronic) for BP1 before cuts. With statistical uncertainties alone, this gives a gain of $1.1$--$1.9$ for BP1--BP3. With 5\% systematics, the gain is $\simeq7$--$8$ for BP1--BP3 in both modes, since the significance is then controlled by $(S/B)/\epsilon$. BP4 and BP5 stay below $1.4\sigma$ in this case, so they require background normalizations controlled at the percent level. The H20 values with systematics assume uncorrelated uncertainties between beam settings and are therefore optimistic; a joint fit with common nuisance parameters is needed.
	
	The selection can still be improved. The common $\slashed{E}$ thresholds correspond to $M_{\rm rec}\gtrsim134$ (240~GeV) and 117~GeV (365~GeV) at the FCC-ee and remove the 90~GeV resonance of BP1, and $\slashed{E}>500$~GeV at the ILC is essentially the kinematic minimum for a $Z$ candidate; a benchmark-dependent $M_{\rm rec}$ window or a shape fit would help, in particular for BP1, BP4 and BP5. The hadronic results should also be cross-checked with initial-state-radiation-enabled $e^+e^-\to q\bar q(\gamma)$ samples. Natural extensions are scans of the $(M_V,g_v)$ and $(M_V,M_{H_2})$ planes overlaid on the relic-density, direct-detection constraints, angular observables of the $Z$ decay products, and the complementary mono-photon, mono-$W$ and vector-boson fusion channels.	
	
	\section*{Acknowledgments}
	R.R. is supported in part by the National Research Foundation of Korea (NRF) Grant No. RS-2026-25484206, and the BK-21 FOUR program through NRF. S. P. would like to thank CERN TH-Division for their support and kind hospitality during his stay at CERN. In the analysis and preparation of this manuscript, Claude and ChatGPT were used to assist with literature searches, refining the language, modifying the LaTeX code, discussing and clarifying selected physics questions, and troubleshooting issues in MadGraph. The authors take full responsibility for the content of the manuscript.
	\newpage
	\appendix
	\section{Kinematic Distributions and Cut-Flow Tables for Optimally Polarized Beams}
	\label{app:optimal_beams}
	
\begin{figure}[t!]
		\centering
		\subfloat[M$_{ll}$]{\includegraphics[width=0.48\linewidth]{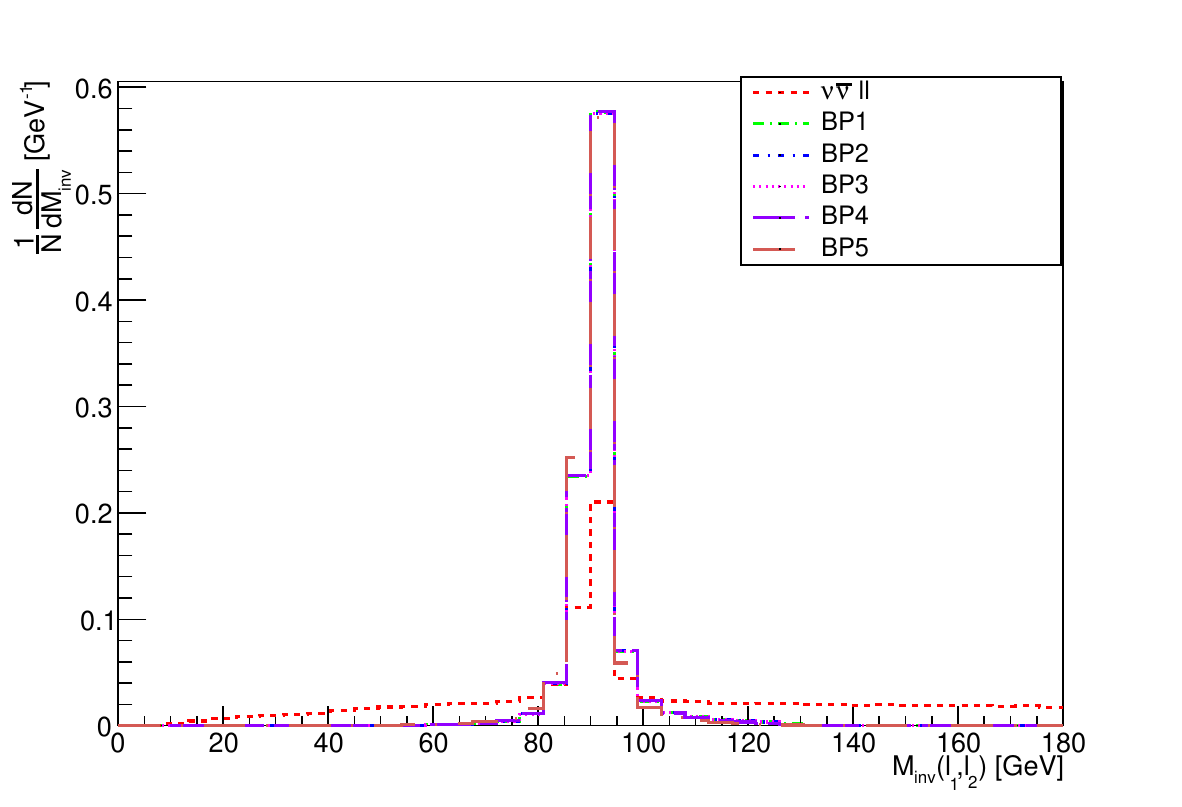}}
		\subfloat[$\slashed{E}_T$]{\includegraphics[width=0.48\linewidth]{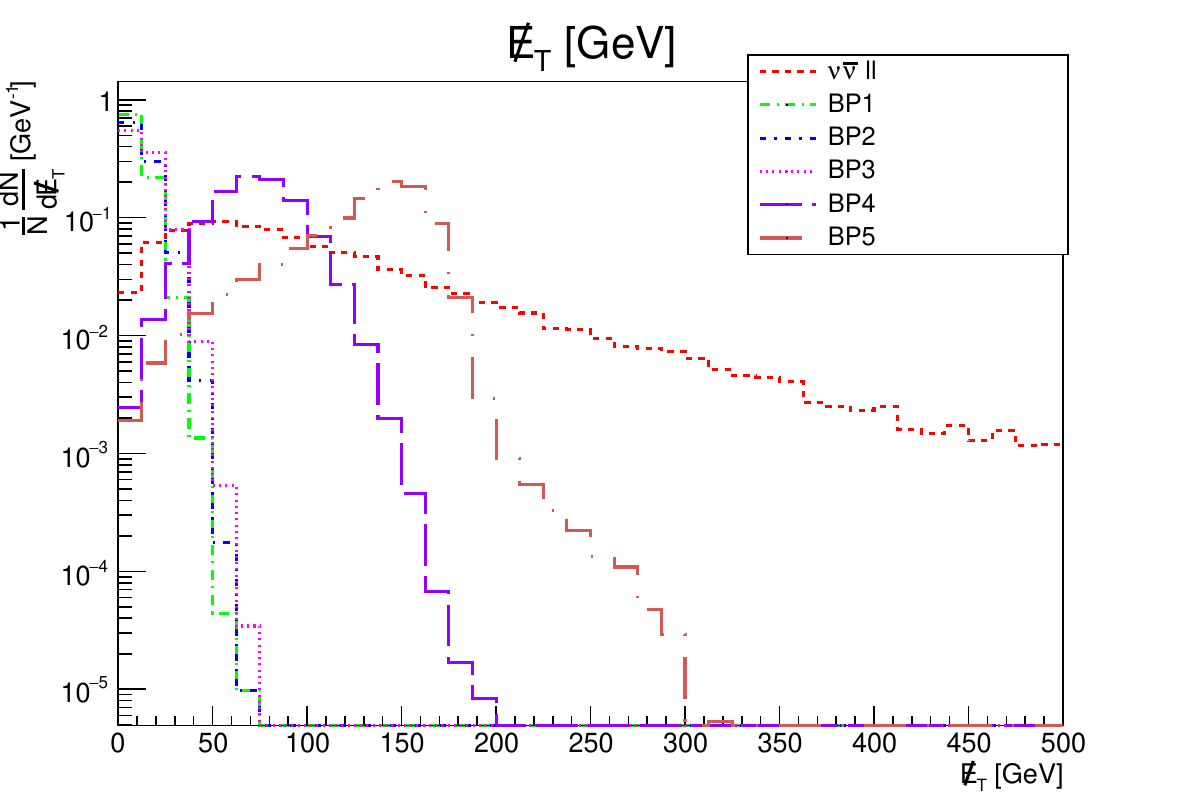}}\\
		\subfloat[$\slashed{E}$]{\includegraphics[width=0.48\linewidth]{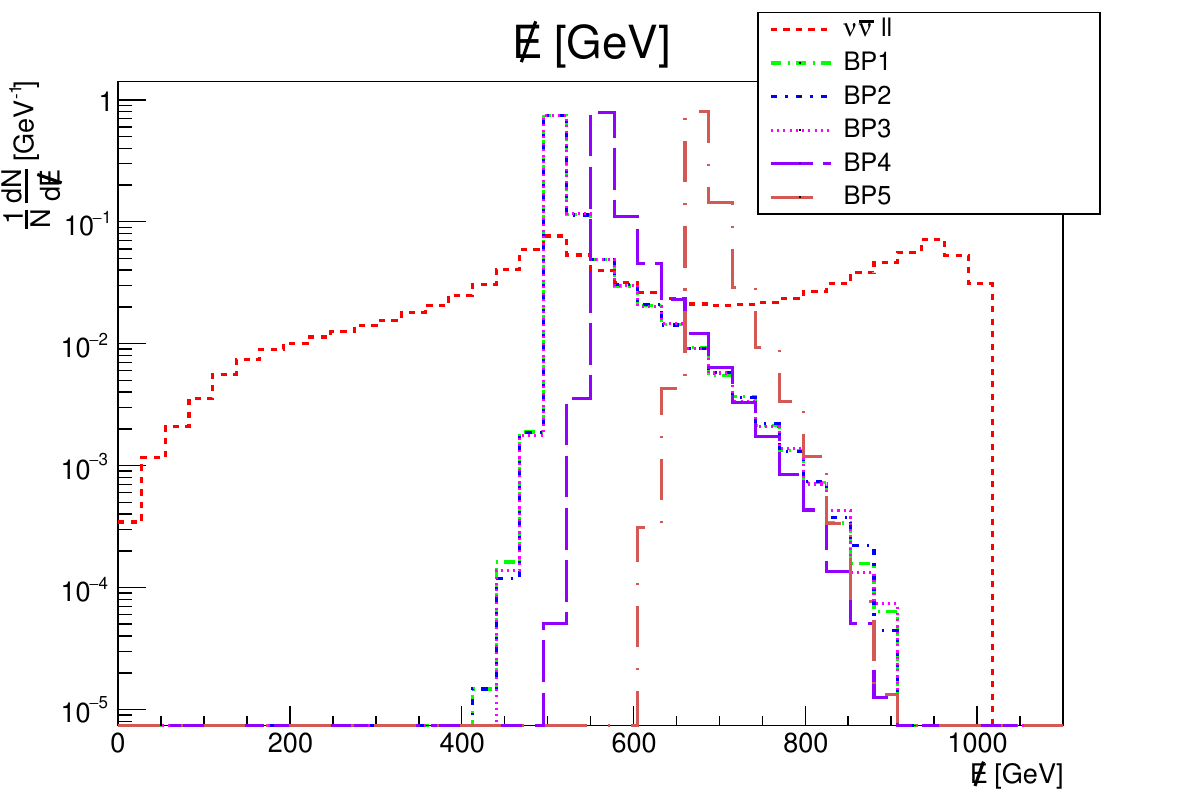}}
		\caption{Simulated signal and background events for optimal beam $(P_{e^-},P_{e^+})=(+80\%,-30\%)$ leptonic mono-Z production at $\sqrt{s}$= 1 TeV ILC.}
		\label{fig:ILC_opt_leptonic}
	\end{figure}

\begin{figure}[h!]
		\centering
		\subfloat[M$_{ll}$]{\includegraphics[width=0.48\linewidth]{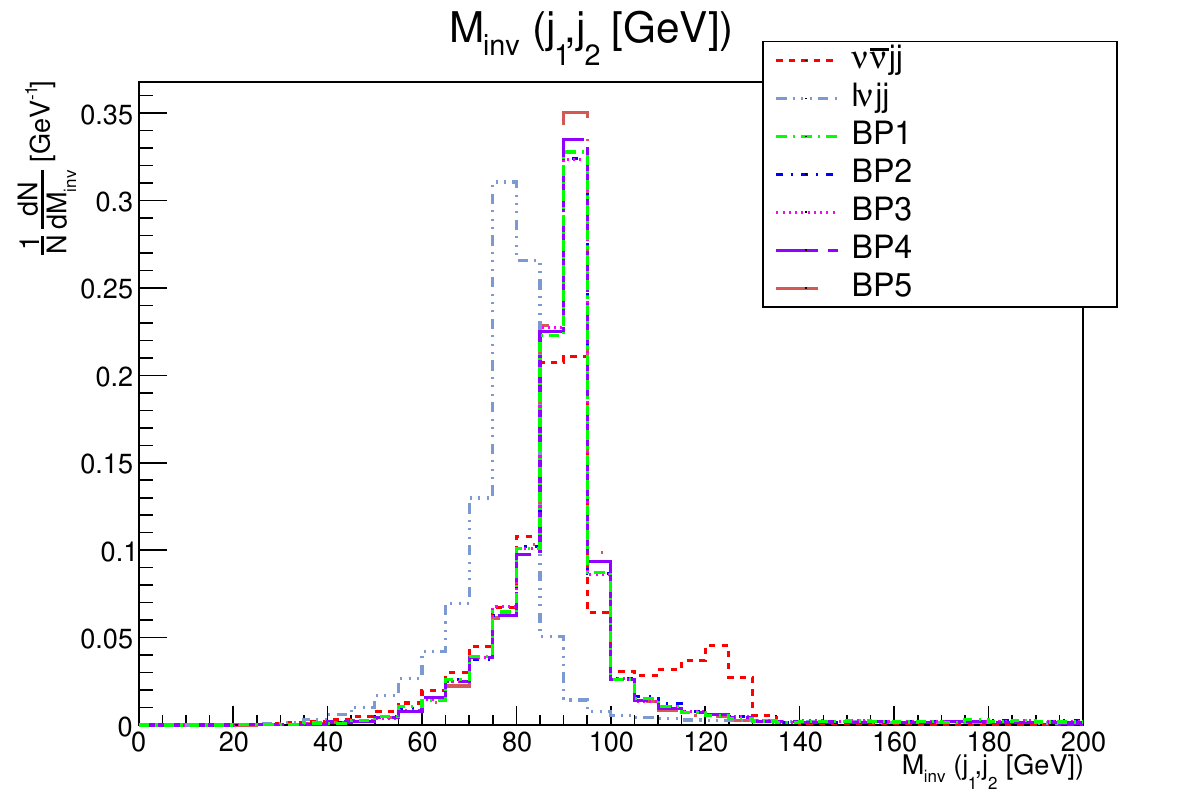}}
		\subfloat[$\slashed{E}_T$]{\includegraphics[width=0.48\linewidth]{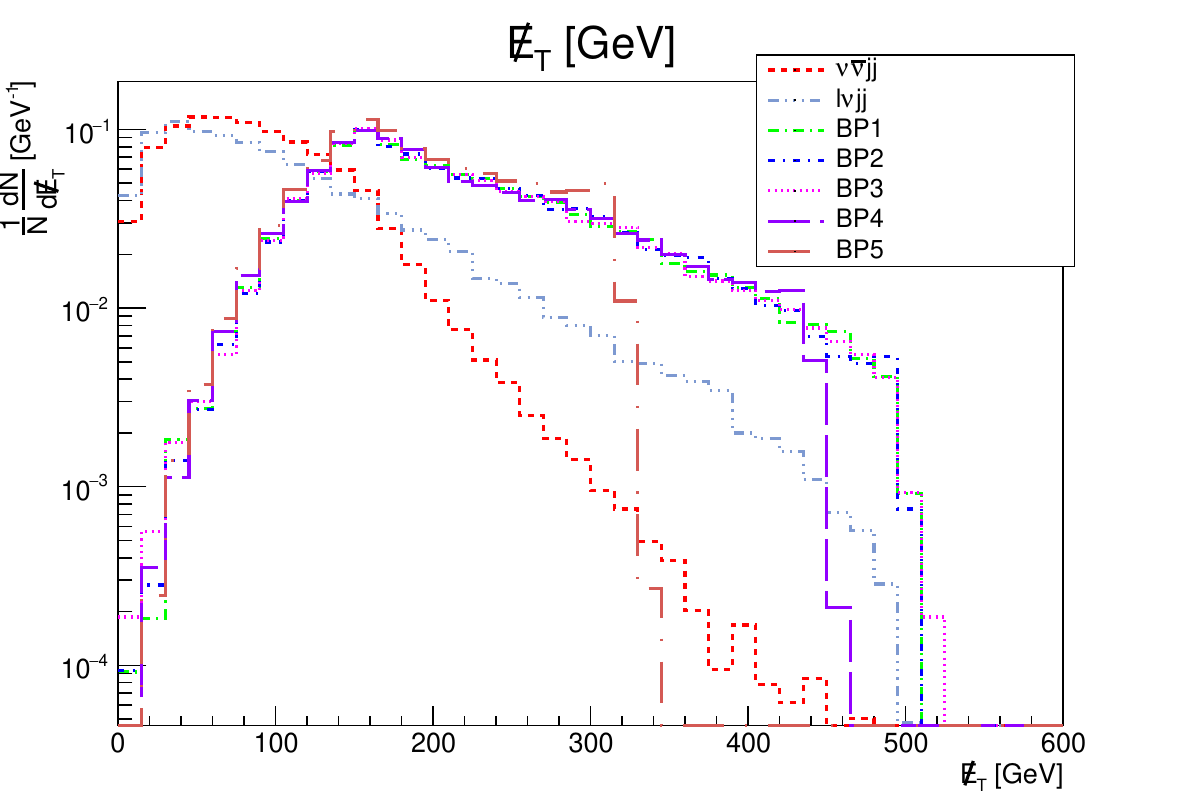}}\\
		\subfloat[$\slashed{E}$]{\includegraphics[width=0.48\linewidth]{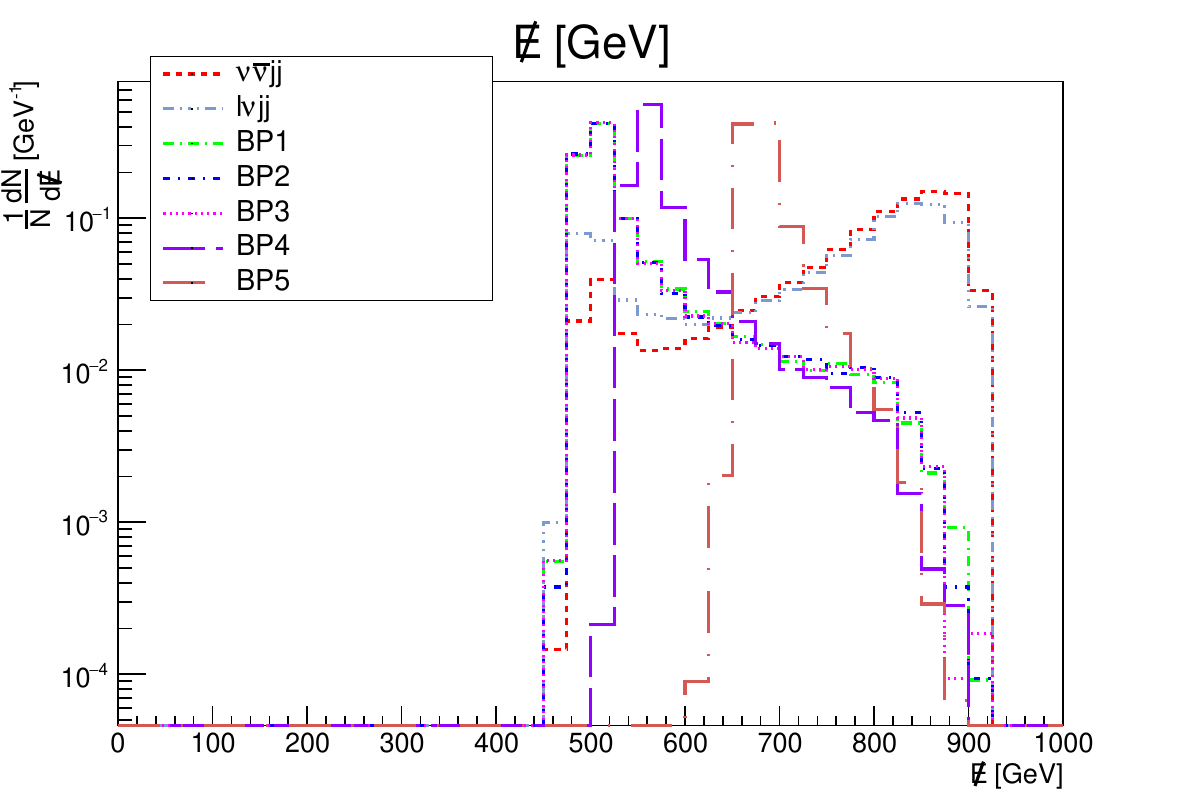}}
		\caption{Simulated signal and background events for optimal beam $(P_{e^-},P_{e^+})=(+80\%,-30\%)$ hadronic mono-Z production at $\sqrt{s}$= 1 TeV ILC.}
		\label{fig:ILC_opt_hadronic}
	\end{figure}

	For completeness, we present the kinematic distributions for the optimally polarized beam configuration, $(P_{e^-},P_{e^+})=(+80\%,-30\%)$, at the ILC with
	$\sqrt{s}=1$ TeV. We consider the same benchmark points and selection cuts as those discussed for the unpolarized beams in the main text.
	In Fig.~\ref{fig:ILC_opt_leptonic}, we show the distributions of the dilepton invariant mass,
	missing transverse energy, and missing energy for the leptonic mono-$Z$ channel. The corresponding distributions for the hadronic mono-$Z$ channel,
	including the dijet invariant mass, missing transverse energy, and missing energy, are shown in Fig.~\ref{fig:ILC_opt_hadronic}.
	The cut-flow tables for the leptonic and hadronic channels are presented in Tabs.~\ref{tab:ILC_opt_leptonic} and
	\ref{tab:ILC_opt_hadronic}, respectively. These tables summarize the number of signal and SM background events after the selection cuts,
	together with their corresponding cut efficiencies, for the optimally polarized beam configuration.
	
	\begin{table}
		\centering
		\caption{Cut-flow table for different BPs for the leptonic mono-$Z$ signal and background at $\sqrt{s}=1$ TeV ILC, considering optimally polarized beams $(P_{e^-},P_{e^+})=(+80\%,-30\%)$.}
		\label{tab:ILC_opt_leptonic}
		\resizebox{\textwidth}{!}{
			\begin{tabular}{lcccccc}
				\hline\hline
				& \multicolumn{1}{c}{SM Background}
				& \multicolumn{5}{c}{Signal} \\
				\cline{2-2} \cline{3-7}
				Selection cuts & $\nu\bar{\nu}\ell^+\ell^-$
				& BP1 & BP2 & BP3 & BP4 & BP5 \\
				\hline
				Baseline Selection
				& 499187 & 499830 & 499828 & 499844 & 499992 & 500000 \\
				& (100\%) & (100\%) & (100\%) & (100\%) & (100\%) & (100\%) \\
				
				After final cut
				& 17580 & 462 & 1790 & 3026 & 4880 & 8479 \\
				& (3.52\%) & (0.09\%) & (0.36\%) & (0.61\%) & (0.98\%) & (1.70\%) \\
				\hline\hline
			\end{tabular}
		}
	\end{table}
	
	\begin{table}
		\centering
		\caption{Cut-flow table for different BPs for the hadronic mono-$Z$ signal and background at $\sqrt{s}=1$ TeV ILC, considering optimally polarized beams $(P_{e^-},P_{e^+})=(+80\%,-30\%)$.}
		\label{tab:ILC_opt_hadronic}
		\resizebox{\textwidth}{!}{
			\begin{tabular}{lccccccc}
				\hline\hline
				& \multicolumn{2}{c}{SM Backgrounds}
				& \multicolumn{5}{c}{Signal} \\
				\cline{2-3} \cline{4-8}
				Selection cuts & $\nu\bar{\nu}jj$ & $\ell\nu jj$
				& BP1 & BP2 & BP3 & BP4 & BP5 \\
				\hline
				Baseline Selection
				& 499977 & 500000 & 500000 & 500000 & 500000 & 500000 & 500000 \\
				& (100\%) & (100\%) & (100\%) & (100\%) & (100\%) & (100\%) & (100\%) \\
				
				After final cut
				& 165572 & 18142 & 8057 & 7836 & 7922 & 14204 & 44815 \\
				& (33.12\%) & (3.63\%) & (1.61\%) & (1.57\%) & (1.58\%) & (2.84\%) & (8.96\%) \\
				\hline\hline
			\end{tabular}
		}
	\end{table}
	
	\clearpage

	\bibliography{References}
	\bibliographystyle{JHEP}
\end{document}